\documentclass{elsart}

 \usepackage{graphics}
\usepackage{epsfig}
\include{epsfig}
\usepackage{amssymb}

\usepackage{amsmath}
\usepackage{graphicx}
\usepackage{psfrag}
\usepackage{dashrule}
\usepackage{amsbsy}
\usepackage{epstopdf}
\usepackage{tikz}
\usetikzlibrary{decorations.markings}
\usetikzlibrary{matrix,shapes,arrows,positioning,chains,shapes.geometric,calc,patterns,decorations.pathmorphing,decorations.markings}
\usepackage{subcaption}
\usepackage{amsfonts}
\usepackage{setspace}
\usepackage{color}
\usepackage{subfloat}
\usepackage{float}
\usepackage{caption}
\usepackage{tabularx}
\usepackage{epsfig}
\usepackage{bm}
\usepackage{dcolumn} 
\usepackage{subfig}
\usepackage{pdfpages}
\graphicspath{{Pics/}} 

\makeatletter
\long\def\@maketablecaption#1#2{\@tablecaptionsize
    \global \@minipagefalse
    \hbox to \hsize{\parbox[t]{\hsize}{\centering #1 \\ #2}}}

\makeatother

\def\vec#1{\mbox{\boldmath $#1$}}
\newcommand{\redcolor}[1]{\textcolor{red}{#1}}
\newcommand{\bluecolor}[1]{\textcolor{blue}{#1}}

\def\G{\Gamma}

\def\Otf{\Omega^\mathrm{f}(t)}

\def\Otnm1f{\Omega^\mathrm{f}(t^{\mathrm{n}-1})}
\def\Otnm12f{\Omega^\mathrm{f}(t^{\mathrm{n}-\frac{1}{2}})}

\def\Otnm1s{\Omega^\mathrm{s}_\mathrm{i}(t^{\mathrm{n}-1})}
\def\Otnm12s{\Omega^\mathrm{s}_\mathrm{i}(t^{\mathrm{n}-\frac{1}{2}})}

\def\usnp1{{\vec u}^\mathrm{s,n+1}}

\def\vphinp1{\vec{\eta}^\mathrm{s,n+1}}

\def\Os{\Omega^\mathrm{s}}
\def\b1{\mbox{\boldmath $1$}}

\def\G{\Gamma}

\def\Otf{\Omega^\mathrm{f}(t)}

\def\Otnm1f{\Omega^\mathrm{f}(t^{\mathrm{n}-1})}
\def\Otnm12f{\Omega^\mathrm{f}(t^{\mathrm{n}-\frac{1}{2}})}

\def\Of{\Omega^\mathrm{f}}
\def\Os{\Omega^\mathrm{s}}
\def\Otnm1s{\Omega^\mathrm{s}(t^{\mathrm{n}-1})}
\def\Otnm12s{\Omega^\mathrm{s}(t^{\mathrm{n}-\frac{1}{2}})}

\newcommand{\nwc}{\newcommand}

\nwc{\qref}[1]{(\ref{#1})} 

\nwc{\ip}[1]{\langle #1 \rangle}

\newcommand{\xx}{\mbox{\boldmath $x$}}

\newcommand{\grad}{\nabla}

\nwc{\ta}{\tilde{a}}

\newtheorem{remark}{Remark}

\def\G{\Gamma}

\def\Otf{\Omega^\mathrm{f}(t)}

\def\Otnm1f{\Omega^\mathrm{f}(t^{\mathrm{n}-1})}
\def\Otnm12f{\Omega^\mathrm{f}(t^{\mathrm{n}-\frac{1}{2}})}

\def\Otnm1s{\Omega^\mathrm{s}(t^{\mathrm{n}-1})}
\def\Otnm12s{\Omega^\mathrm{s}(t^{\mathrm{n}-\frac{1}{2}})}

\begin{document}
\begin{frontmatter}

\title{A Variational Projection Scheme for Nonmatching Surface-to-Line Coupling 
between 3D Flexible Multibody System and Incompressible Turbulent Flow} 

\author[label1,label2]{P. S. Gurugubelli},
\author[label1,label2]{R. Ghoshal},
\author[label1,label2]{V. Joshi} and
\author[label1,label2]{R. K. Jaiman\corauthref{cor}}
\ead{mperkj@nus.edu.sg}
\address[label1]{Department of Mechanical Engineering, National University of Singapore, Singapore}
\address[label2]{Keppel-NUS Corporate Laboratory, Singapore}
\corauth[cor]{Corresponding author. Tel.: + 65 6601 2547; fax: +65 6779 1459.}

\begin{abstract}
This paper is concerned with the partitioned iterative formulation
to simulate the fluid-structure interaction of a nonlinear multibody system 
in an incompressible turbulent flow. The proposed formulation relies on a three-dimensional (3D) incompressible turbulent flow solver, a nonlinear monolithic elastic structural solver for 
constrained flexible multibody system  and the  nonlinear iterative force correction scheme 
for coupling of the turbulent fluid-flexible multibody system with nonmatching interface meshes. 
While the fluid equations are discretized using a stabilized Petrov-Galerkin formulation in space and the generalized-$\alpha$ updates in time, the multibody system utilizes a discontinuous space-time Galerkin finite element method. 
We address two key challenges in the present formulation. Firstly, the coupling of the incompressible turbulent flow with a system of nonlinear elastic bodies described in a co-rotated frame. 
Secondly, the projection of the tractions and displacements 
across the nonmatching 3D fluid surface elements and the one-dimensional 
line elements for the flexible multibody system in a conservative manner.
Through the nonlinear iterative correction and the conservative projection, 
the developed fluid-flexible multibody interaction solver is stable for problems involving 
strong inertial effects between the fluid-flexible multibody system and the coupled interactions 
among each multibody component. 
The accuracy of the proposed coupled finite element framework is validated against the available experimental 
data for a long flexible cylinder undergoing vortex-induced vibration in a uniform current flow condition. 
Finally, a practical application of the proposed framework is demonstrated by 
simulating the flow-induced vibration of a realistic offshore 
floating platform connected to a long riser and an elastic mooring system.
\end{abstract}
\end{frontmatter}
\section{Introduction}
\begin{figure}
    \centering
    \includegraphics [width=0.98\textwidth]{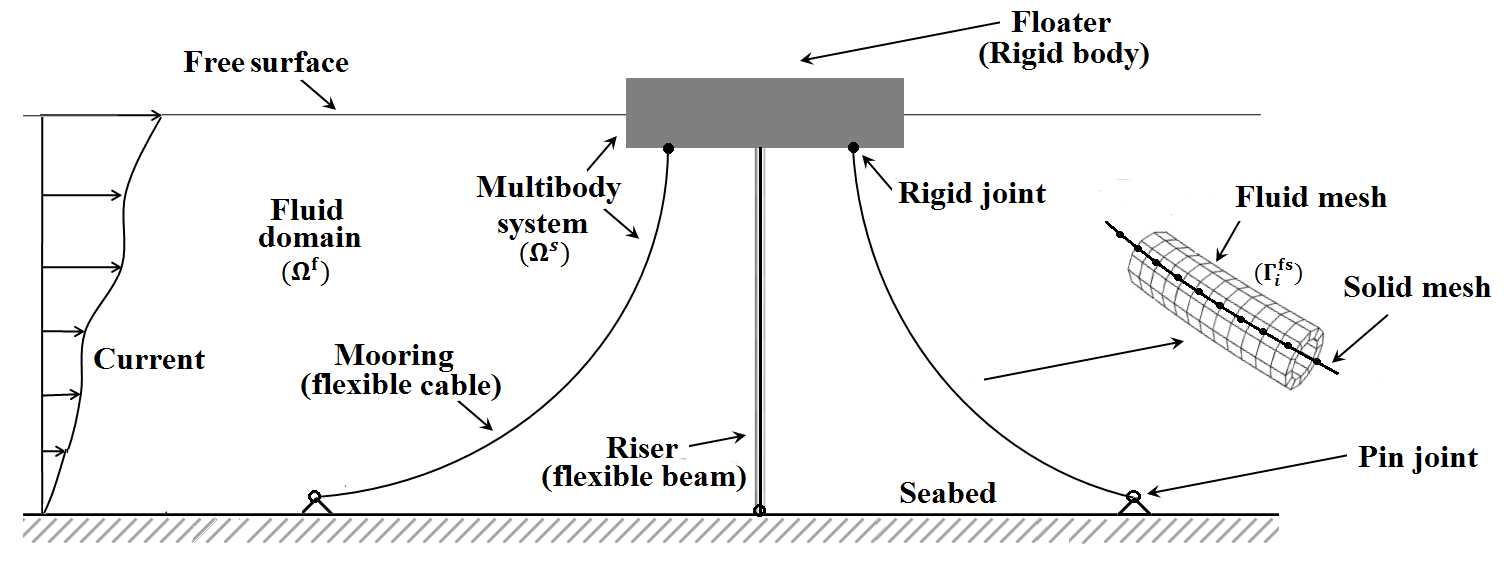}
    \caption{Schematic of a typical floater-mooring system consisting of a rigid floater body with a long flexible riser and moorings which is exposed to high velocity ocean currents and free surface effects.}
    \label{multibody_interface}
\end{figure}

The interaction between multiple interconnected rigid or flexible bodies with the surrounding fluid flow is ubiquitous in engineering applications ranging from underwater robotics, bio-inspired structures, helicopter rotor dynamics to offshore wind turbines and oil/gas platforms. Such fluid-flexible multibody interactions (FFMI) are typically characterized by large rigid body displacement, rotation and local deformation of the flexible structure due to nonlinear fluid dynamic forces along the interface. Through a strong coupled interaction, such large displacements or deformations in turn alter the flow field around the multibody system and the fluid loading acting back on them. Such strong coupled dynamical interaction has influence on the performance of structural material and the stability of flexible multibody system.
In particular, offshore/ocean engineering applications involve complex interactions between multiple flexible structures such as  marine risers and mooring lines with strong underwater currents leading to vortex-induced-vibration (VIV).  These flexible structures are typically connected to rigid bodies such as a drill-ship or a floater which interacts with ocean waves and currents. 
While a flexible marine  riser is used to transport hydrocarbon from the subsea well-head on the ocean floor to the floating structure, the mooring lines are used for the station-keeping of floating offshore structure. A typical schematic for the floater-mooring system is illustrated in Fig. \ref{multibody_interface}, whereas the multibody system is exposed to ocean 
current and free surface effects. 
The prediction and control of the complex interaction between a floating body and flexible multibody structures are 
crucial for the offshore industry. There have been extensive experimental and semi-empirical research works in the past for this practical coupled dynamical problem of vessel-riser-mooring. However, there are not many studies focusing on the fully-coupled analysis of the flexible multibody system in a realistic ocean environment. 
The development of coupled variational formulation for a floater-mooring-riser system poses numerous difficulties due to strong coupling of ocean current flow with the floater,
the riser and the mooring lines.

Numerical simulations of FFMI are generally accomplished by using either partitioned or monolithic schemes \cite{turek2006,bazilevs2008,gee2011,jaiman_ficf2015,jaiman_CF16}. In a monolithic approach, fluid and structural equations are assembled into a single block and solved as a unified entity. Monolithic schemes are robust but they lack the advantage of flexibility and modularity of using existing stable and well established fluid and/or structural solvers \cite{turek2006,bazilevs2008,gee2011,jaiman_ficf2015,jaiman_CF16}. In order to overcome such difficulties, partitioned approach is popularly used in which the fluid and the structural equations are solved in a sequential manner by satisfying the velocity (Dirichlet) and traction (Neumann) continuity  along the interface to achieve the desired stability and accuracy in the coupling \cite{Felippa2001,Jaiman_2011a,Jaiman_2011b}. In addition to the traditional monolithic and partitioned methods, one can also have a new class of coupling techniques where the subdomains are selectively decoupled from monolithic framework without losing desired features such as numerical stability and computation cost. One such method has been proposed in \cite{jaiman_ficf2015} wherein the authors have decoupled the fluid mesh motion and structural positions from the monolithic framework thereby improving the computational cost. 

In the literature, several partitioned methods can be found that can deal with the fluid-structure coupling. These methods can be broadly classified into strongly-coupled and loosely-coupled. In strongly coupled (implicit) methods \cite{Matthies_2006,Ahn_2006,Zhang2004,Dettmer_2006,jaiman_CMAME2016,jaiman_CF16}, predictor-corrector type of iterations are performed at each time step to ensure the convergence of the interface properties. However, it is well known in the field of offshore engineering \cite{jaiman_eccomas} that strongly-coupled partitioned iterative schemes suffer from convergence related issues due to predominant added mass effects \cite{blom_2015cmame,van2009added,Forster_2007,nobile,dettmer2007,heil2008,jaiman_CMAME2016}. In loosely-coupled (explicit) methods \cite{Felippa2001,Guruswamy_1990,cebral1997conservative,farhat05,Jaiman_2011a,Jaiman_2011b}, governing equations of each sub-domains are separately marched in time, satisfying the velocity and traction continuity along the interfaces through algebraic jump conditions in a staggered fashion. These schemes often suffer from numerical instability and temporal inaccuracy caused by the jump along the interface due to the time lag \cite{blom1998,Jaiman_2011a}.

Apart from the numerical instability arising from the relative inertia of immersed solid and displaced fluid, another primary challenge for an FFMI involving multiple structures arises from the significant differences in the relative inertia and/or material properties between each of the structural subdomains \cite{Bauchau_2003}. Similar to the fluid-structure coupling, one can use either monolithic or partitioned techniques for coupling of the multiple interconnected structural domains i.e. flexible multibody systems with constraints. Among the monolithic schemes, significant work has been carried out with energy preserving (EP) schemes and energy decaying (ED) schemes for multibody interactions. For a nonlinear multibody system, EP schemes perform poorly when applied to a system that is physically stiff and ED schemes become necessary \cite{bauchau2010flexible,shabana1998}. Oscillations are particularly violent in multibody simulations due to the presence of algebraic constraints and the nonlinearities of the system provide a mechanism to transfer energy from the low to the high frequency modes. Hence, the presence of high frequency numerical dissipation is an indispensable feature of robust time integrators for a flexible multibody system. To deal with a system where high frequencies are present, i.e., the system is physically stiff, various ED schemes have been proposed \cite{bauchau1999design,Bauchau199637,BOTTASSO1997393}.

Recently, a loosely coupled partitioned staggered technique with improved stability and accuracy has been proposed in \cite{yenduri2017new} to deal with the problem of strong inertial coupling between multibody components. In this method, constraints are transformed into a mathematically equivalent partial differential equation, which includes a coupling parameter for the dimensional consistency of velocity. However, in this method a simplified force decomposition technique has been utilized to estimate the fluid forces, thereby reducing the complexity involving the fluid-structure coupling. This partitioned method has been extended in \cite{Meng201751} for conjugate heat transfer problems based on a discretization of the interface coupling conditions using a generalized Robin (mixed) condition. Even though there have been a lot of efforts to deal with the relative inertia and/or material properties, it is worth mentioning that none of the methods has been proven to be stable for all possible scenarios. Moreover, the stabilization of this scheme often depends on arbitrary constants. Therefore, to avoid any complexity in fluid-flexible multibody coupling an unconditionally stable and accurate monolithic scheme is utilized in the present manuscript \cite{bauchau2007dymore}. For the development of present coupling algorithm, a time discontinuous Galerkin scheme based on energy decay inequality is utilized for multibody interaction where constraints are typically enforced through the Lagrange multiplier technique.

Various types of flexible multibody systems with constraints are widely used in offshore engineering, viz. floater-mooring system, wave energy converter, offshore wind turbines. Precise load and motion control of these systems are often challenging, especially in the harsh environment due to highly nonlinear dynamic loads \cite{Ghoshal_2016,law2017wake,Mysa2016}. Despite known for their shortcomings, present day commercial packages still widely use semi-empirical force decomposition methods such as Morrison's equation to calculate the load on the structure. There are some other studies in which flow-induced vibration of the offshore platforms are studied based on models with rigid body mounted on top of a spring in which an equivalent stiffness is assumed for the mooring system. Only few publications on 3D flexible multibody interaction with an incompressible flow 
can be found in the open literature.
In this study, a fully coupled fluid-structure interaction (FSI) solver is developed to simulate the flow-induced vibration of the multibody system with constraints (viz., floater-mooring-riser system) in a turbulent flow. The structural domain of flexible multibody system is solved via geometrically nonlinear co-rotational finite element method, whereas the fluid domain is solved using Petrov-Galerkin finite element method for moving boundary Navier-Stokes solutions. 
A partitioned iterative scheme is used that relies on the nonlinear iterative force correction (NIFC) \cite{jaiman_CMAME2016,jaiman_CF16}, for the numerical stabilization of the coupling between the incompressible turbulent flow and multibody dynamics. In the NIFC method, the coupled sub-domains are marched in time separately and the interface force correction is constructed at the end of each fluid subiteration. 
For high Reynolds number effects, the flow turbulence is modeled using Spalart-Allmaras (S-A) based Delayed Detached Eddy Simulation (DDES) via a positivity preserving variational (PPV) method \cite{PPV}. To test the accuracy, the proposed variational framework is validated against the full-scale marine riser VIV experiments in a uniform current flow. 
Finally, we demonstrate the the proposed framework to simulate a realistic fully-coupled floater-mooring-riser system in a turbulent current flow.

In the present paper, two challenges related to the variationally  coupled fluid-flexible multibody system are addressed: 
(i) the coupling of incompressible turbulent flow with a collection of flexible and rigid bodies, 
(ii) projection of the forces and motions across three-dimensional (3D) fluid surface elements and one-dimensional (1D) elastic line elements. In this regard, the nonlinear iterative force correction scheme \cite{jaiman_CF16} has been extended  to incorporate a collection of constrained elastic bodies. To enable the interface coupling between the nonmatching 3D fluid mesh and 1D mesh of the multibody components, variables are projected from elastic line elements onto the wetted surface elements and vice-versa. The fluid traction at the quadrature points are projected onto the target elastic line elements using quadrature-projection scheme. On the other hand, the nodal displacement and velocity vectors from 1D multibody line elements are mapped onto the 3D fluid mesh at the interface using nodal-projection. The solution obtained from the solid and fluid solvers are marched in time independently and approximate interface force corrections are evaluated by the generalization of Aitken's $\Delta^2$ extrapolation. This provides a convergent and stable update of the forces at the fluid-structure interface even at a low structure-to-fluid mass ratio. Moreover, a monolithic unconditionally stable energy decaying scheme \cite{Bauchau199637,bauchau1999design} for the flexible multibody system ensures solution convergence of the coupled solver even if the system has large difference in relative inertia and/or material properties between each structural subdomains. Therefore, the proposed scheme addresses the existing challenges in developing a partitioned-staggered solver for FFMI and this warrants to undertake a wide range of problems in engineering applications.

The outline of the rest of the article is as follows. In Section \ref{NM}, governing equations for the turbulent flow (based on the Navier-Stokes and SA-DDES) and the flexible multibody system along with the coupling procedure is presented, wherein line-to-surface coupling between the fluid and the structure for nonmatching meshes is described for the riser and the mooring lines. Variational formulation for the flow, the flexible multibody system with constraints and the PPV discretization of the turbulence transport equation as well as the NIFC-based fluid-flexible multibody coupling technique are covered in Section \ref{Var}. Section \ref{Val} presents the validation of the proposed framework in which simulations of VIV of a flexible riser in uniform current is performed and compared with the available experimental data. In Section \ref{App}, a practical application of the coupled FFMI solver to study the dynamics of coupled floater-riser-mooring is presented. The major conclusions of this work are reported in Section \ref{Conclusion}.
\section{Governing Equations of Fluid and Flexible Multibody System}\label{NM}

In this section, we first present the governing differential equations of the fluid-flexible multibody solver based on the Navier-Stokes and the flexible multibody equations with constraints. Afterwards, the closure problem for turbulence based on the delayed detached eddy simulation is described. Finally, the methodology to treat the fluid-structure interface by maintaining the continuity of velocity and traction along the fluid-structure interface is presented.

\subsection{The incompressible Navier-Stokes equations}
The governing equations for the incompressible fluid are formulated in an arbitrary Lagrangian-Eulerian (ALE) framework. 
The unsteady Reynolds averaged Navier-Stokes equations for an incompressible flow are
\begin{align} 
	\rho^\mathrm{f}\frac{\partial \bar{\boldsymbol{v}}^\mathrm{f}}{\partial t}\bigg|_{\hat{x}^\mathrm{f}} + \rho^\mathrm{f}(\bar{\boldsymbol{v}}^\mathrm{f} - \boldsymbol{v}^\mathrm{m})\cdot\nabla\bar{\boldsymbol{v}}^\mathrm{f} &= \nabla\cdot \bar{\boldsymbol{\sigma}}^\mathrm{f} + \nabla\cdot{\boldsymbol{\sigma}}^\mathrm{des} + \boldsymbol{b}^\mathrm{f}\ \ \ \mathrm{on\ \ \Omega^\mathrm{f}(t)}, \label{eq:NS} \\
	\nabla\cdot\bar{\boldsymbol{v}}^\mathrm{f} &= 0\ \ \ \mathrm{on\ \ \Omega^\mathrm{f}(t)}, \label{eq:continuity}
\end{align}
where $\bar{\boldsymbol{v}}^\mathrm{f}=\bar{\boldsymbol{v}}^\mathrm{f}(\xx^\mathrm{f},t)$ and $\boldsymbol{v}^\mathrm{m}=\boldsymbol{v}^\mathrm{m}(\xx^\mathrm{f},t)$ represent the fluid and mesh velocities defined for each spatial point $\xx^\mathrm{f} \in  \Otf$, respectively. $\boldsymbol{b}^\mathrm{f}$ is the body force applied on the fluid and $\bar{\boldsymbol{\sigma}}^\mathrm{f}$ is the Cauchy stress tensor for a Newtonian fluid, written as
\begin{equation}
\bar{\boldsymbol{\sigma}}^\mathrm{f} = -\bar{p} \boldsymbol{I} + \mu^\mathrm{f}\left(\boldsymbol{\nabla} \bar{\boldsymbol{v}}^\mathrm{f} + \left(\boldsymbol{\nabla} \bar{\boldsymbol{v}}^\mathrm{f}\right)^T\right),
\label{eq:cauchyStress}
\end{equation}
where $\bar{p}$ denotes the time averaged fluid pressure, $\mu^\mathrm{f}$ is the dynamic viscosity of the fluid and $\boldsymbol{\sigma}^\mathrm{des}$ is the turbulent stress term. The spatial and temporal coordinates are denoted by $\xx^\mathrm{f}$ and $t$, respectively. The first term in Eq.~(\ref{eq:NS}) represents the partial derivative of $\bar{\boldsymbol{v}}^\mathrm{f}$ with respect to time with the ALE referential coordinate $\hat{x}^\mathrm{f}$ kept fixed. A multibody system immersed in fluid may undergo deformation and flow-induced vibration due to the unsteady fluid forces. We next present the governing equations of the flexible multibody system, which allow large rotations and large 
displacements, but small deformations of the flexible bodies via co-rotational approach. 

\subsection{Flexible multibody system}
The equation of motion for a flexible structure $\Omega^\mathrm{s}_i$ with Lagrangian material points $\boldsymbol{X}^\mathrm{s}$ in curvilinear coordinate system reads as
\begin{equation}
\rho^\mathrm{s}\frac{\partial^2 {\boldsymbol{u}}^\mathrm{s}}{\partial t^2}(\boldsymbol{X}^\mathrm{s},t)+\nabla \cdot \vec{\sigma}^\mathrm{s}\left(\boldsymbol{E}\left(\boldsymbol{u}^\mathrm{s}\right)\right)=\boldsymbol{b}^\mathrm{s} \qquad \forall \boldsymbol{X}^\mathrm{s}\in \Omega^\mathrm{s}_i,\label{eq:structural}
\end{equation}
where $\boldsymbol{u}^\mathrm{s}$ represents the structural displacement, $\rho^\mathrm{s}$ is the structural density, $\vec{\sigma}^\mathrm{s}$ is the first Piola-Kirchhoff stress tensor, $\boldsymbol{E}\left(\boldsymbol{u}^\mathrm{s}\right)= 1/2\left[\left(\boldsymbol{I}+\nabla\boldsymbol{u}^\mathrm{s}\right)^T\left(\boldsymbol{I}+\nabla\boldsymbol{u}^\mathrm{s}\right)-\boldsymbol{I}\right]$ denotes the Cauchy-Green Lagrangian strain tensor and $\boldsymbol{b}^\mathrm{s}$ is the body force acting on the multibody $\Omega_i^\mathrm{s}$ and $i$ is the $i^\mathrm{th}$ structural component of the multibody system.
Here, the body velocity is the measured from an inertial frame at a point of reference configuration and the body displacement $\boldsymbol{u}^\mathrm{s}$ is a nonlinear function of the unknown rigid body displacements and flexible body deformations. We employ the Lagrangian formulation  and assume linear material behavior.
Traditionally, the Cauchy-Green strain tensor for a geometrically linear formulation neglects the higher order terms in $\vec{\grad}\vec{u}^\mathrm{s}$. However, such simplification of strain cannot describe large rigid-body deformations. Hence, we decompose the structural displacement $\vec{u}^\mathrm{s}$ as the sum of large rigid body displacements ($\boldsymbol{u}_0^\mathrm{s}$) and small deformation ($\tilde{\boldsymbol{u}}^\mathrm{s}$), i.e., $\boldsymbol{u}^\mathrm{s}=\boldsymbol{u}_0^\mathrm{s}+\tilde{\boldsymbol{u}}^\mathrm{s}$. The rigid body displacement is given as $\boldsymbol{u}_0^\mathrm{s} = \boldsymbol{u}_R^\mathrm{s} +\boldsymbol{R}\boldsymbol{X}^\mathrm{s} - \boldsymbol{X}^\mathrm{s}$, where $\boldsymbol{u}_R^\mathrm{s}$ represents the rigid body displacement and $\boldsymbol{R}$ is the conformal rotation matrix chosen to parametrize the finite rotation. The rotation matrix can also be expressed as $\boldsymbol{R}=\vec{\nabla}\vec{u}_0+\vec{I}$. Therefore, the Cauchy-Green Lagrangian strain tensor can be rewritten as
\begin{equation}
	\vec{E}=\frac{1}{2}\left[\left(\vec{R}+\vec{\nabla}\tilde{\boldsymbol{u}}^\mathrm{s}\right)\left(\vec{R}+\vec{\nabla}\tilde{\boldsymbol{u}}^\mathrm{s}\right)-\vec{I}\right].
	\end{equation}
Neglecting the quadratic terms in $\vec{\grad}\tilde{\boldsymbol{u}}^\mathrm{s}$, the strain tensor can be simplified as 
\begin{equation}
\tilde{\vec{E}}=\frac{1}{2}\left(\left(\vec{\nabla}\tilde{\vec{u}}^\mathrm{s}\right)^T\vec{R}+\vec{R}^T\tilde{\vec{u}}^\mathrm{s}\right).
\end{equation}

We will briefly present the kinematic constraints on the relative motion of various bodies of the multibody system in Section \ref{Var}, such that general dynamics of 
flexible multibody systems can be effectively simulated. The work done by the associated forces due to the constraints must vanish during the relative motion of connected bodies.
In the present formulation,  the kinematic constraints are implemented via Lagrange multipliers, for more details refer to \cite{bauchau2010flexible}.
\subsection{Treatment of the fluid-structure interface}
Coupling of the fluid and the multibody system consisting of multiple interconnected components such as beam, cable and rigid body is carried out by satisfying the continuity of velocity and traction along the fluid-flexible multibody interface of each component. Let the fluid-flexible multibody interface for the $i^\mathrm{th}$ component at $t=0$ be denoted as
$\G_i^\mathrm{fs}=\Of(0) \cap \Os_i$ and the interface at time $t$ as $\G_i^\mathrm{fs}(t)=\boldsymbol{\varphi}^\mathrm{s}(\G_i^\mathrm{fs},t)$. Here $\boldsymbol{\varphi}^\mathrm{s}$ represents the position vector which maps the initial position $\boldsymbol{X}^\mathrm{s}$ of the flexible multibody to its position at time $t$, i.e., $\boldsymbol{\varphi}^\mathrm{s}(\boldsymbol{X}^\mathrm{s},t) = \boldsymbol{X}^\mathrm{s} + \boldsymbol{u}^\mathrm{s}(\boldsymbol{X}^\mathrm{s},t)$. The velocity and the traction continuity at each interface $\Gamma_i^\mathrm{fs}$ can be written as
\begin{align}
\vec{\bar{v}}^\mathrm{f}\left(\boldsymbol{\varphi}^\mathrm{s}(\boldsymbol{X}^\mathrm{s},t),t\right) &= \vec{v}^\mathrm{s}\left(\boldsymbol{X}^\mathrm{s},t\right)\qquad  &\forall \boldsymbol{X}\in \Gamma^\mathrm{fs}_i, 
\label{eq:bcsVelocity} \\
\int_{\boldsymbol{\varphi}^\mathrm{s}(\gamma,t)} \bar{\boldsymbol{\sigma}}^\mathrm{f}\left(\boldsymbol{x}^\mathrm{f},t\right)\cdot\boldsymbol{\mathrm{n}} \mathrm{d} \Gamma 
+ \int_{\gamma} \vec{t}^\mathrm{s} \mathrm{d} \Gamma &=0\qquad &\forall \gamma \in \Gamma^\mathrm{fs}_i ,
\label{eq:bcsTraction}
\end{align}
where $\vec{v}^\mathrm{s}$ is the structural velocity at time $t$ defined as $\vec{v}^\mathrm{s} = \partial\boldsymbol{\varphi}^\mathrm{s}/\partial t$, $\boldsymbol{\mathrm{n}}$  is the outer normal to the fluid-structure interface, $\gamma$ is any part of the fluid-flexible multibody interface ($\Gamma^\mathrm{fs}_i$) in the reference configuration and ${\boldsymbol{\varphi}^\mathrm{s}(\gamma,t)}$ is the corresponding fluid part at time $t$.

\subsection{Closure problem for flow turbulence: Delayed detached eddy simulation}
For the sake of completeness, we briefly present the closure problem for the turbulence modeling of the Navier-Stokes equations for high Reynolds number flows. The turbulent stress term in Eq.~(\ref{eq:NS}) is modeled using the Boussinesq approximation,
\begin{align}
	\boldsymbol{\sigma}^\mathrm{des} = \mu^\mathrm{f}_T\left(\boldsymbol{\nabla} \bar{\boldsymbol{v}}^\mathrm{f} + \left(\boldsymbol{\nabla} \bar{\boldsymbol{v}}^\mathrm{f}\right)^T\right),
\end{align}
where $\mu_T^\mathrm{f}$ is the turbulent dynamic viscosity given by $\mu_T^\mathrm{f} = \nu_T \rho^\mathrm{f}$. Here, $\nu_T$ is the turbulent kinematic viscosity. $\nu_T$ is related to the eddy viscosity $\tilde{\nu}$ by $\nu_T = \tilde{\nu}f_{v1}$, where
\begin{align}
	f_{v1} = \frac{\tilde{\chi}^3}{\tilde{\chi}^3 + c_{v1}^3},\quad \tilde{\chi} = \frac{\tilde{\nu}}{\nu}.
\end{align}
$\nu$ is the molecular viscosity given as $\nu = \mu^\mathrm{f}/\rho^\mathrm{f}$ and $\tilde{\nu}$ is solved by the transport equation
\begin{equation}
	\frac{\partial\tilde{\nu}}{\partial t} + (\bar{\vec{v}}^\mathrm{f} - \vec{v}^\mathrm{m})\cdot\nabla \tilde{\nu} = c_{b1}\tilde{S}\tilde{\nu} + \frac{1}{\sigma}\nabla\cdot\big[ (\nu +\tilde{\nu} ) \nabla \tilde{\nu} \big] + \frac{c_{b2}}{\sigma}(\nabla\tilde{\nu})\cdot(\nabla\tilde{\nu}) - c_{w1}f_w\bigg[ \frac{\tilde{\nu}}{\tilde{d}} \bigg]^2
\end{equation}
where $\tilde{S}=S+(\tilde{\nu}/(\kappa^2 \tilde{d}^2))f_{v2}$, $S$ being the magnitude of vorticity. $c_{b1}$, $c_{b2}$, $\sigma$, $\kappa$, $c_{w1}$ and $c_{v1}$ are constants defined for the Spalart-Allmaras model in \cite{Spalart_1}. The distance $\tilde{d}$ from the wall is defined in such a way that the model acts in RANS mode in the attached boundary layer region and switches to LES mode in the separated flow region, thus providing the advantages of both reduction in computational cost and accuracy in the separated regions. More details can be found in \cite{Jochen,Spalart_2}.

\section{Variational Formulation of Fluid-Flexible Multibody Interaction}\label{Var}

\subsection{Flow solver}
We present the stabilized Petrov-Galerkin variational form of the Navier-Stokes equations in this section. We employ the generalized-$\alpha$ variational time integration technique \cite{Gen_alpha} to march the variables in time which can be unconditionally stable and second-order accurate for linear problems. The generalized-$\alpha$ method for fluid flow describes
\begin{align}
\vec{\bar{v}}^\mathrm{f, n+\alpha}  = \alpha^\mathrm{f} \vec{\bar{v}}^\mathrm{f,n+1} &+ \left(1-\alpha^\mathrm{f}\right)\vec{\bar{v}}^\mathrm{f, n}, \quad
\partial_t\vec{\bar{v}}^\mathrm{f, n+\alpha_\mathrm{m}}  = \alpha_\mathrm{m}^\mathrm{f} \partial_t\vec{\bar{v}}^\mathrm{f, n+1}  + (1-\alpha_\mathrm{m}^\mathrm{f}) \partial_t\vec{\bar{v}}^\mathrm{f, n} \notag \\
& \mbox{and} \quad \vec{v}^\mathrm{m, n+\alpha}  = \alpha^\mathrm{f} \vec{v}^\mathrm{m,n+1} +\left(1-\alpha^\mathrm{f}\right)\vec{v}^\mathrm{m,n}, \label{alphaF}
\end{align}
where 
$
\vec{\bar{v}}^\mathrm{f, n+1}  = \vec{\bar{v}}^\mathrm{f, n}  + \Delta t\left(\left(1-\gamma^\mathrm{f}\right) \partial_t\vec{\bar{v}}^\mathrm{f, n} +\gamma^\mathrm{f} \partial_t\vec{\bar{v}}^\mathrm{f, n+1} \right)$,
$\alpha^\mathrm{f},\ \alpha_\mathrm{m}^\mathrm{f}$ and 
$\gamma^\mathrm{f}$ are the fluid solver integration parameters as described in \cite{Dettmer_2006,yuri,jaiman_ficf2015}.

Let the domain $\Omega^\mathrm{f}$ be discretized into $\mathrm{n}_\mathrm{el}^\mathrm{f}$ number of three-dimensional Lagrange finite elements such that $\Omega^\mathrm{f} = \cup_{e=1}^\mathrm{n_{el}} \Omega^e$ and $\emptyset = \cap_{e=1}^\mathrm{n_{el}} \Omega^e$. Consider $\mathcal{S}^\mathrm{f,h}$ as the space of trial solution which satisfy the Dirichlet boundary condition and $\mathcal{V}^\mathrm{f,h}$ as the space of test function which is null on the Dirichlet boundary. The variational statement for the flow equations in Eqs.~(\ref{eq:NS} and \ref{eq:continuity}) using a Petrov-Galerkin framework is:
find $[\bar{\vec{v}}_\mathrm{h}^\mathrm{f,n+\alpha^f}, \bar{p}_\mathrm{h}^\mathrm{f,n+1}]\in \mathcal{S}^\mathrm{f,h}$ such that $\forall [\boldsymbol{\phi}^\mathrm{f}_\mathrm{h}, q_\mathrm{h}] \in \mathcal{V}^\mathrm{f,h}$:
\begin{align} \label{weakForm_NS}
&\int_{\Omega^\mathrm{e}} \rho^\mathrm{f} ( \partial_t\bar{\boldsymbol{v}}_\mathrm{h}^\mathrm{f,n+\alpha_m^f} + (\bar{\boldsymbol{v}}_\mathrm{h}^\mathrm{f,n+\alpha^f} - \boldsymbol{v}_\mathrm{h}^\mathrm{{m},n+\alpha^f}) \cdot \nabla\bar{\boldsymbol{v}}_\mathrm{h}^\mathrm{f,n+\alpha^f})\cdot \boldsymbol{\phi}_\mathrm{h}^\mathrm{f} \mathrm{d\Omega} \nonumber \\
&+ \int_{\Omega^\mathrm{e}} \bar{\boldsymbol{\sigma}}_\mathrm{h}^\mathrm{f,n+\alpha^f} : \nabla \boldsymbol{\phi}_\mathrm{h}^\mathrm{f} \mathrm{d\Omega} + \int_{\Omega^\mathrm{e}} {\boldsymbol{\sigma}^\mathrm{des}}_\mathrm{h}^\mathrm{f,n+\alpha^f} : \nabla \boldsymbol{\phi}_\mathrm{h}^\mathrm{f} \mathrm{d\Omega} \nonumber \\
&- \int_{\Omega^\mathrm{e}} \nabla q_\mathrm{h}\cdot \bar{\boldsymbol{v}}_\mathrm{h}^\mathrm{f,n+\alpha^f} \mathrm{d\Omega}\nonumber \\
&+ \displaystyle\sum_\mathrm{e=1}^\mathrm{n_{el}^\mathrm{f}}\int_{\Omega^\mathrm{e}} \tau_\mathrm{m} (\rho^\mathrm{f}(\bar{\boldsymbol{v}}_\mathrm{h}^\mathrm{f,n+\alpha^f} - {\boldsymbol{v}}_\mathrm{h}^\mathrm{{m},n+\alpha^f})\cdot \nabla\boldsymbol{\phi}_\mathrm{h}^\mathrm{f} + \nabla q_\mathrm{h} )\cdot \boldsymbol{\mathcal{R}}_\mathrm{m}(\bar{\boldsymbol{v}}^\mathrm{f},\bar{p}) \mathrm{d\Omega^e}\nonumber \\
&+ \displaystyle\sum_\mathrm{e=1}^\mathrm{n_{el}^\mathrm{f}}\int_{\Omega^\mathrm{e}} \nabla\cdot \boldsymbol{\phi}_\mathrm{h}^\mathrm{f}\tau_\mathrm{c} \nabla\cdot \bar{\boldsymbol{v}}_\mathrm{h}^\mathrm{f,n+\alpha^f} \mathrm{d\Omega^e}\nonumber \\
& -\displaystyle\sum_\mathrm{e=1}^\mathrm{n_{el}^\mathrm{f}}\int_{\Omega^\mathrm{e}} \tau_\mathrm{m} \boldsymbol{\phi}_\mathrm{h}^\mathrm{f}\cdot (\boldsymbol{\mathcal{R}}_\mathrm{m}(\bar{\boldsymbol{v}}^\mathrm{f},\bar{p}) \cdot \nabla \bar{\boldsymbol{v}}_\mathrm{h}^\mathrm{f,n+\alpha^f}) \mathrm{d\Omega^e}\nonumber \\
&-\displaystyle\sum_\mathrm{e=1}^\mathrm{n_{el}^\mathrm{f}}\int_{\Omega^\mathrm{e}} \nabla \boldsymbol{\phi}_\mathrm{h}^\mathrm{f}:(\tau_\mathrm{m}\boldsymbol{\mathcal{R}}_\mathrm{m}(\bar{\boldsymbol{v}}^\mathrm{f},\bar{p}) \otimes \tau_\mathrm{m}\boldsymbol{\mathcal{R}}_\mathrm{m}(\bar{\boldsymbol{v}}^\mathrm{f},\bar{p})) \mathrm{d\Omega^e}\nonumber \\
&= \int_{\Omega^\mathrm{e}} \boldsymbol{b}^\mathrm{f}(t^\mathrm{n+\alpha^f})\cdot \boldsymbol{\phi}_\mathrm{h}^\mathrm{f} \mathrm{d\Omega} + \int_{\Gamma_\mathrm{h}} \boldsymbol{h}^\mathrm{f}\cdot \boldsymbol{\phi}^\mathrm{f}_\mathrm{h} \mathrm{d\Gamma}.
\end{align}
where $\boldsymbol{\phi}^\mathrm{f}_\mathrm{h}$ and $q_\mathrm{h}$ represent the test functions for the fluid velocity and pressure respectively. The first, second and third lines in the above Eq.~(\ref{weakForm_NS}) represent the Galerkin terms for the Navier-Stokes equations, the fourth and fifth lines the Petrov-Galerkin stabilization terms for the momentum and continuity equations respectively, and sixth and seventh lines denote the approximation of the fine scale velocity on the element interiors based on the multi-scale argument \cite{Hughes_conserve,Hsu,Akkerman}. $\boldsymbol{\mathcal{R}}_\mathrm{m}(\bar{\boldsymbol{v}}^\mathrm{f},\bar{p})$ is the residual of the momentum equation at the element level and $\tau_\mathrm{m}$ and $\tau_\mathrm{c}$ are the stabilization parameters added to the element level integrals \cite{Brooks,Hughes_X,Tezduyar_1,France_II}. The details of the definitions of the stabilization parameters can be found in \cite{Vaibhav_CAF}.

\subsection{Turbulence solver}
To maintain the consistency of time integration, the turbulence transport equation is also discretized in time using the generalized-$\alpha$ method. The transport equation to be solved for closure problem can be written in the form of a convection-diffusion-reaction equation as follows
	\begin{align} \label{CDR}
	\frac{\partial\tilde{\nu}}{\partial t} + \mathbf{u}\cdot\nabla\tilde{\nu} - \nabla\cdot(k\nabla\tilde{\nu}) + r\tilde{\nu} = 0,
	\end{align}
	where
	\begin{align}
	\mathbf{u} =\ (\bar{\boldsymbol{v}}^\mathrm{f} - \boldsymbol{v}^\mathrm{m}) - \frac{c_{b2}}{\sigma}\nabla\tilde{\nu}, \qquad
	k = \frac{\nu + \tilde{\nu}}{\sigma}, \qquad
	r = c_{w1}f_{w}\frac{\tilde{\nu}}{\tilde{d}^2} - c_{b1}\tilde{S}.
	\end{align}
	Similar to the variational form of the flow system, the variational statement for the turbulence system can be written as
	find $\tilde{\nu}_\mathrm{h}^\mathrm{f,n+\alpha^f} \in\mathcal{S}^\mathrm{f,h}$ such that $\forall \psi^\mathrm{f}_\mathrm{h} \in \mathcal{V}^\mathrm{f,h}$:
	\begin{align}
	\label{sa_ddes_var}
	&\int_{\Omega^\mathrm{e}} ( \partial_t \tilde{\nu}_\mathrm{h}^\mathrm{f,n+\alpha_m^f} + \mathbf{u}\cdot \nabla\tilde{\nu}_\mathrm{h}^\mathrm{f,n+\alpha^f} + r\tilde{\nu}_\mathrm{h}^\mathrm{f,n+\alpha^f}) \psi_\mathrm{h}^\mathrm{f} \mathrm{d\Omega} \nonumber \\
	& + \int_{\Omega^\mathrm{e}} k \nabla \tilde{\nu}_\mathrm{h}^\mathrm{f,n+\alpha^f}\cdot \nabla\psi_\mathrm{h}^\mathrm{f} \mathrm{d\Omega} \nonumber \\
	& + \displaystyle\sum_\mathrm{e=1}^\mathrm{n_{el}}\int_{\Omega^\mathrm{e}} ( \mathbf{u}\cdot \nabla \psi_\mathrm{h}^\mathrm{f} + |r|\psi_\mathrm{h}^\mathrm{f} )\tau_\mathrm{t} \mathcal{R}_\mathrm{t}(\tilde{\nu}) \mathrm{d\Omega^\mathrm{e}}\nonumber \\
	& + \displaystyle\sum_\mathrm{e=1}^\mathrm{n_{el}}\int_{\Omega^\mathrm{e}} \chi \frac{|\mathcal{R}_\mathrm{t}(\tilde{\nu})|}{|\nabla \tilde{\nu}_\mathrm{h}^\mathrm{f,n+\alpha^f}|} k_\mathrm{s}^\mathrm{add} \nabla\psi_\mathrm{h}^\mathrm{f}\cdot \bigg( \frac{\mathbf{u} \otimes \mathbf{u}}{|\mathbf{u}|^2} \bigg) \cdot \nabla \tilde{\nu}_\mathrm{h}^\mathrm{f,n+\alpha^f} \mathrm{d\Omega^\mathrm{e}} \nonumber \\
	& + \displaystyle\sum_\mathrm{e=1}^\mathrm{n_{el}}\int_{\Omega^\mathrm{e}} \chi \frac{|\mathcal{R}_\mathrm{t}(\tilde{\nu})|}{|\nabla \tilde{\nu}_\mathrm{h}^\mathrm{f,n+\alpha^f}|} k_\mathrm{c}^\mathrm{add} \nabla\psi_\mathrm{h}^\mathrm{f}\cdot \bigg( \mathbf{I} - \frac{\mathbf{u} \otimes \mathbf{u}}{|\mathbf{u}|^2} \bigg) \cdot \nabla \tilde{\nu}_\mathrm{h}^\mathrm{f,n+\alpha^f} \mathrm{d\Omega^\mathrm{e}} \nonumber \\
	&= \int_{\Gamma_\mathrm{h}} \psi_\mathrm{h}^\mathrm{f} g \mathrm{d\Gamma}, 
	\end{align}
	where the first and second line represents the Galerkin terms, the third line is the linear stabilization term and the fourth and fifth lines correspond to the nonlinear stabilization terms which impart the positivity preserving property to the solution. $\mathcal{R}_\mathrm{t}$ is the residual of the transport equation and $\tau_\mathrm{t}$ is the stabilization parameter.  The details about the positivity preserving variational formulation with the definition of the parameters $\chi$, $k^\mathrm{add}_s$ and $k^\mathrm{add}_c$ can be found in \cite{PPV,Vaibhav_CAF}.

\subsection{Multibody solver with constraints}
Let the multibody domain $\Omega^\mathrm{s}_i\subset \mathbb{R}^d$, where $d=3$, be discretized into $n_\mathrm{el}^\mathrm{s}$ 1D line elements. 
Discretization of the multibody components into 1D line elements give an advantage of the reduction in the computational cost for large scale structural systems which are encountered in the offshore applications. 
Let us consider $\mathcal{S}^\mathrm{s,h}$ as the trial function space and $\mathcal{V}^\mathrm{s,h}$ as the test function space which is null on the Dirichlet boundary. Weak variational form of the multibody system in Eq.~(\ref{eq:structural}) can be written using the principle of virtual work as
\begin{align}
	\int_{t^n}^{t^{n+1}}\left(\int_{\Omega^\mathrm{s}_i}\rho^\mathrm{s}\frac{\partial^2 {\boldsymbol{u}}^\mathrm{s}_\mathrm{h}}{\partial t^2}\cdot \vec{\phi}^\mathrm{s}_\mathrm{h}\ \mathrm{d}\Omega+ \int_{\Omega^\mathrm{s}_i} \vec{\sigma}^\mathrm{s}\left(\tilde{\boldsymbol{E}}\left(\boldsymbol{u}^\mathrm{s}_\mathrm{h}\right)\right): \vec{\grad}\vec{\phi}^\mathrm{s}_\mathrm{h}\ \mathrm{d}\Omega\right) \ \mathrm{d}t = \notag \\ \int_{t^n}^{t^{n+1}}\left(\int_{\Omega^\mathrm{s}_i}\boldsymbol{b}^\mathrm{s}\cdot \vec{\phi}^\mathrm{s}_\mathrm{h}\ \mathrm{d}\Omega+\int_{\Gamma_i}\vec{t}^\mathrm{s}\cdot \vec{\phi}^\mathrm{s}_\mathrm{h}\ \mathrm{d}\Gamma\right)\ \mathrm{d}t \label{eq:FMD_var},
	\end{align}
where $\vec{\phi}^\mathrm{s}_\mathrm{h}$ denotes the test function for the structural displacements and the term $\boldsymbol{t}^\mathrm{s}$ represents the fluid tractions acting along the interface $\Gamma_i$ between the fluid $\Omega^\mathrm{f}$ and the multibody $\Omega^\mathrm{s}_i$. A detailed derivation of the above weak form can be found in \cite{bauchau2010flexible}.
The kinematic joints or connections that restrict the  motion of the interconnected multibodies are generally described using a constraint equation which is given as
\begin{equation}
\boldsymbol{c}(\boldsymbol{u}^\mathrm{s})=0.
\end{equation}
A penalty method is employed in the present formulation to model the constraint 
and this yields the governing equation for the multibody system as   
\begin{align}
		&\int_{t^n}^{t^{n+1}}\left(\int_{\Omega^\mathrm{s}_i}\rho^\mathrm{s}\frac{\partial^2 {\boldsymbol{u}}^\mathrm{s}_\mathrm{h}}{\partial t^2}\cdot \vec{\phi}^\mathrm{s}_\mathrm{h}\ \mathrm{d}\Omega+ \int_{\Omega^\mathrm{s}_i} \vec{\sigma}^\mathrm{s}\left(\tilde{\boldsymbol{E}}\left(\boldsymbol{u}^\mathrm{s}_\mathrm{h}\right)\right): \vec{\grad}\vec{\phi}^\mathrm{s}_\mathrm{h}\ \mathrm{d}\Omega +\int_{\Gamma_i}\boldsymbol{c}'(\boldsymbol{u}^\mathrm{s}_\mathrm{h})^T\lambda^h \cdot \vec{\phi}^\mathrm{s}_\mathrm{h} \ \mathrm{d}\Gamma \right) \ \mathrm{d}t \notag \\ = &\int_{t^n}^{t^{n+1}}\left(\int_{\Omega^\mathrm{s}_i}\boldsymbol{b}^\mathrm{s}\cdot \vec{\phi}^\mathrm{s}_\mathrm{h}\ \mathrm{d}\Omega+\int_{\Gamma_i} \boldsymbol{t}\cdot \vec{\phi}^\mathrm{s}_\mathrm{h}\ \mathrm{d}\Omega\right)\ \mathrm{d}t \label{weakFormMBD},\\
	&\boldsymbol{c}(\boldsymbol{u}^\mathrm{s}_\mathrm{h})=0\label{eq:FMD_Const},
	\end{align}
where $\lambda^h$ is the Lagrange multiplier corresponding to the constraints Eq.~(\ref{eq:FMD_Const}) and $\boldsymbol{c}'$ denotes the Jacobian of $\boldsymbol{c}$. We can rewrite the above variational form of the multibody system with constraints in Eq.~(\ref{weakFormMBD}) in a much simplified matrix form as
\begin{equation}
\int_{t^n}^{t^{n+1}}\left(\vec{M}\ddot{\vec{u}^\mathrm{s}}(t)+\vec{K}\vec{u}^\mathrm{s}(t)+\vec{C}\vec{u}^\mathrm{s}(t) \right)\ \mathrm{d}t= \int_{t^n}^{t^{n+1}}\left(\vec{F}^\mathrm{s}(t)\right)\ \mathrm{d}t,\label{weakFormMbd}
\end{equation} 
where $\vec{M}$, $\vec{K}$ and $\vec{C}$ denote the mass, stiffness and constraint matrices of the multibody system. $\boldsymbol{F}^\mathrm{s}$ comprises of both the body force and the external fluid forces acting on the multibody. 
The above constraints are discretized in such a manner the constraint forces do not produce any work at the discrete solution level.

\begin{remark} Depending on the type of component of the structure, e.g., beam, cable, etc., the construction of these matrices will be different. Detailed derivation with regard to each specific element under co-rotational finite element framework can be found in \cite{Bauchau_2003}.
\end{remark}

An unconditionally stable energy decaying scheme is applied to update structural variables temporally. The scheme is obtained by applying a linear time discontinuous Galerkin approximation to the flexible multibody equation (Eq.~(\ref{weakFormMbd})) between the initial ($t^\mathrm{n}$) and the final time ($t^\mathrm{n+1}$). A linear approximation of the Lagrange multiplier ($\lambda^h$) is carried out over the time step, $t \in [t^\mathrm{n}, t^\mathrm{n+1}]$. The resulting discretized equations of motion are as follows:
\begin{align} \label{cable_r}
\bm{M}\frac{\dot{\bm{u}}_\mathrm{n+1}^\mathrm{s,-}-\dot{\bm{u}}_\mathrm{n}^\mathrm{s,-}}{\Delta t^\mathrm{s}}+\bm{K}\frac{\dot{\bm{u}}_\mathrm{n+1}^\mathrm{s,-} + \dot{\bm{u}}_\mathrm{n}^\mathrm{s,+}}{2} + \frac{\lambda^{h,-}_\mathrm{n+1} + \lambda^{h,+}_\mathrm{n}}{2} \frac{\vec{C}\left({\bm{u}}_\mathrm{n+1}^\mathrm{s,-}\right) + \vec{C}\left({\bm{u}}_\mathrm{n}^\mathrm{s,-}\right)}{2} &= \frac{\dot{\boldsymbol{F}}_\mathrm{n}^\mathrm{s,+} + \dot{\boldsymbol{F}}_\mathrm{n+1}^\mathrm{s,-}}{2}, \\
\bm{M}\frac{\dot{\bm{u}}_\mathrm{n}^\mathrm{s,+} - \dot{\bm{u}}_\mathrm{n}^\mathrm{s,-}}{\Delta t^\mathrm{s}}-\bm{K}\frac{\dot{\bm{u}}_\mathrm{n+1}^\mathrm{s,-} - \dot{\bm{u}}_\mathrm{n}^\mathrm{s,+}}{6} + \frac{\lambda^{h,-}_\mathrm{n+1}-\lambda^{h,+}_\mathrm{n}}{6} \frac{\vec{C}\left({\bm{u}}_\mathrm{n}^\mathrm{s,+}\right) + \vec{C}\left({\bm{u}}_\mathrm{n}^\mathrm{s,-}\right)}{2} &=\frac{\dot{\boldsymbol{F}}_\mathrm{n}^\mathrm{s,+} - \dot{\boldsymbol{F}}_\mathrm{n+1}^\mathrm{s,-}}{6}, \\
\frac{{\bm{u}}_\mathrm{n+1}^\mathrm{s,-} - {\bm{u}}_\mathrm{n}^\mathrm{s,-}}{\Delta t^\mathrm{s}} = \frac{\dot{\bm{u}}_\mathrm{n+1}^\mathrm{s,-} + \dot{\bm{u}}_\mathrm{n}^\mathrm{s,+}}{2}, \\
\frac{{\bm{u}}_\mathrm{n}^\mathrm{s,+} - {\bm{u}}_\mathrm{n}^\mathrm{s,-}}{\Delta t^\mathrm{s}} = -\frac{1}{6}\left[\dot{\bm{u}}_\mathrm{n+1}^\mathrm{s,-} - \dot{{\bm{u}}}_\mathrm{n}^\mathrm{s,-} - \alpha\left(\dot{\bm{u}}_\mathrm{n}^\mathrm{s,+} - \dot{{\bm{u}}}_\mathrm{n}^\mathrm{s,-}\right)\right],\label{cable_r4}
\end{align}
where $\Delta t^\mathrm{s}$ is the time step size for the structural system and $\alpha$ is a tuning parameter that controls the amount of numerical dissipation by the scheme. $\alpha=1$ is chosen for the present study which ensures asymptotic annihilation. The notations $()^-_\mathrm{n}$, $()^+_\mathrm{n}$ and $()^-_\mathrm{n+1}$ used in Eqs.~(\ref{cable_r}-\ref{cable_r4}) indicate the corresponding quantities at $t^-_\mathrm{n}$, $t^+_\mathrm{n}$ and $t^-_\mathrm{n+1}$, respectively. Following from the theory of the time discontinuous Galerkin method applied to hyperbolic conservation laws, this scheme can be proven to be unconditionally stable based 
on energy decay inequality \cite{Bauchau199637,bauchau1999design,Bauchau_2003}.

\begin{remark} The generalized-$\alpha$ method is employed for the fluid system to integrate in time between $t \in [t^\mathrm{n}, t^\mathrm{n+1}]$, which is unconditionally stable and second-order accurate for linear problems, whereas an energy decaying scheme proposed in \cite{Bauchau199637} is utilized to solve the multibody structural dynamics. Since energy conservation is not sufficient to yield a robust time integration scheme, high frequency numerical dissipation must be added in nonlinear flexible multibody system.
\end{remark} 

\subsection{Fluid-flexible multibody interface}

The coupling between the nonmatching 3D fluid mesh elements and the 1D multibody line elements is carried out through conservative surface-to-line coupling and vice-versa. Transfer of the structural displacements onto the fluid mesh is carried out by the line-to-surface coupling while the fluid forces are transferred onto the structure via the surface-to-line coupling. We present the description of these coupling procedure in this section.

\subsubsection{Line-to-surface coupling}
\label{l_to_s_coupling}
\begin{figure}
	\centering
	\begin{subfigure}{.58\textwidth}
		\centering
		\includegraphics[width=0.99\columnwidth]{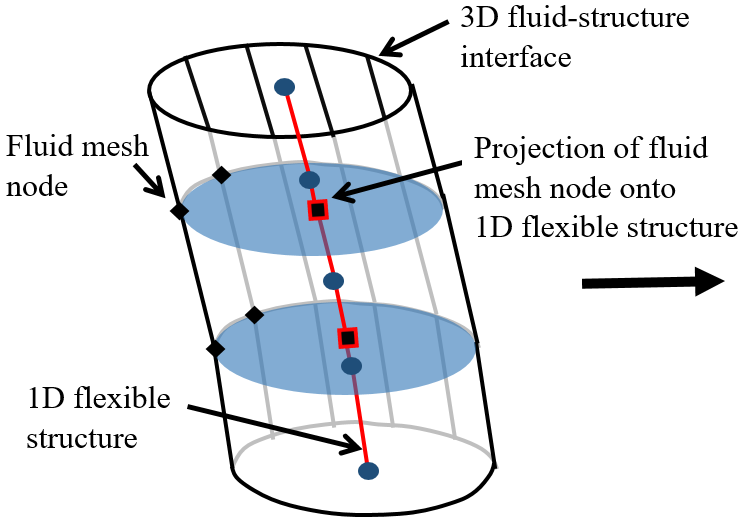}
		\caption{}
	\end{subfigure}%
	\begin{subfigure}{.41\textwidth}
		\centering
		\includegraphics[width=0.99\columnwidth]{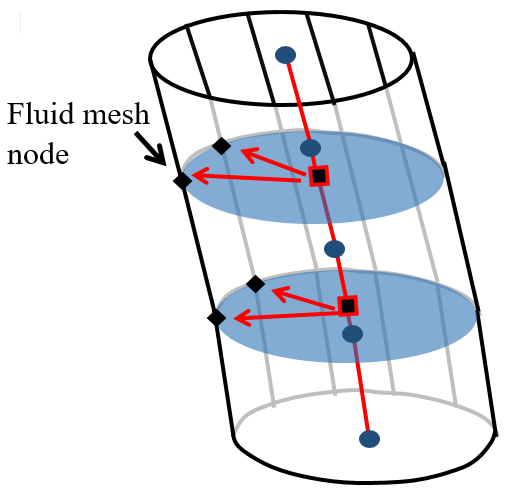}
		\caption{}
	\end{subfigure}
	\caption{Schematic of line-to-surface projection of displacements and velocity: (a) search for the corresponding structural line node for the fluid mesh nodes and (b) projection of the variables to the fluid mesh nodes.}
	\label{fig:l-to-s}
\end{figure}

This section briefly describes the transfer of the structural displacements to the fluid side while satisfying the ALE compatibility and the velocity continuity condition at the fluid-structure interface $\Gamma^\mathrm{fs}_i$. A nodal projection scheme is used to transfer the nodal displacements and velocity of each multibody component onto the targeted fluid surface mesh (as depicted in Fig. \ref{fig:l-to-s}).
As mentioned earlier, we need to satisfy the ALE compatibility and the velocity continuity while transferring the displacements. This is carried out in the following manner:
We first find the projection of the fluid mesh nodes on the one-dimensional structural line element (see Fig. \ref{fig:l-to-s}a). The displacements and velocities at the projected point on the line element are then interpolated using the structural shape functions. The interpolated values of the variables at the projected structural nodes are then assigned to the corresponding fluid node:
\begin{align}
\boldsymbol{d}^\mathrm{m,n+1}_k\left(\vec{x}^\mathrm{f,n+1}_k\right) =\sum_{j=1}^{ns} \boldsymbol{\phi}_j^\mathrm{s}\left(\vec{\mathbb{P}}\left(\vec{x}^\mathrm{f,n+1}_k\right)\right) \left(\boldsymbol{u}^\mathrm{s,-}_{n+1}\right)_j\quad \forall \vec{x}^\mathrm{f,n+1}_k\in\Gamma^\mathrm{fs}_i(t^\mathrm{n+1}),
\end{align}
where $\boldsymbol{d}^\mathrm{m,n+1}_k$ and $\vec{x}^\mathrm{f,n+1}_k$ represent the fluid mesh 
displacement and position of the $k^\mathrm{th}$ node at time $t^\mathrm{n+1}$ respectively,  the function $\vec{\mathbb{P}}\left(\vec{x}^\mathrm{f,n+1}_k\right)$ is the projection function which projects the fluid mesh node $k \in \Gamma^\mathrm{fs}_i(t^\mathrm{n+1})$ onto the structural line elements, $\vec{\phi}^\mathrm{s}_j$ denotes the structural shape function for the 1D flexible multibody mesh node $j$ and 
$ns$ is the number of 1D structural mesh nodes. The velocity continuity at the interface is then satisfied by equating the fluid velocity with the mesh velocity for all the fluid nodes located on the interface $\Gamma^\mathrm{fs}_i(t^\mathrm{n+\alpha^\mathrm{f}})$, i.e.
	\begin{align}
	\bar{\boldsymbol{v}}^\mathrm{f,n+\alpha^f}_k= \boldsymbol{v}^\mathrm{m,n+\alpha^f}_k= \frac{\boldsymbol{d}^\mathrm{m,n+1}_k - \boldsymbol{d}^\mathrm{m,n}_k}{\Delta t}.
	\end{align}
	This completes the description of the line-to-surface coupling to transfer the displacements from the structure to the fluid side.

\subsubsection{Surface-to-line coupling}
\label{s_to_l_coupling}
\begin{figure}
	\centering
	\begin{subfigure}{.55\textwidth}
		\centering
		\includegraphics[width=0.99\columnwidth]{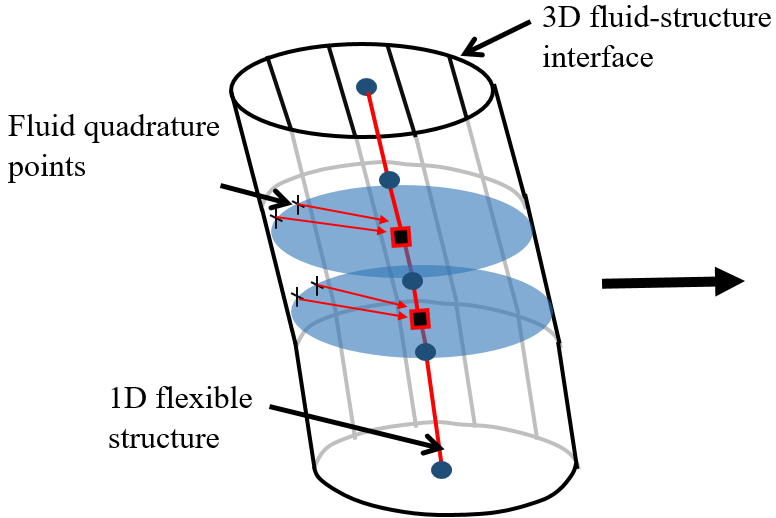}
		\caption{}
	\end{subfigure}%
	\begin{subfigure}{.44\textwidth}
		\centering
		\includegraphics[width=0.99\columnwidth]{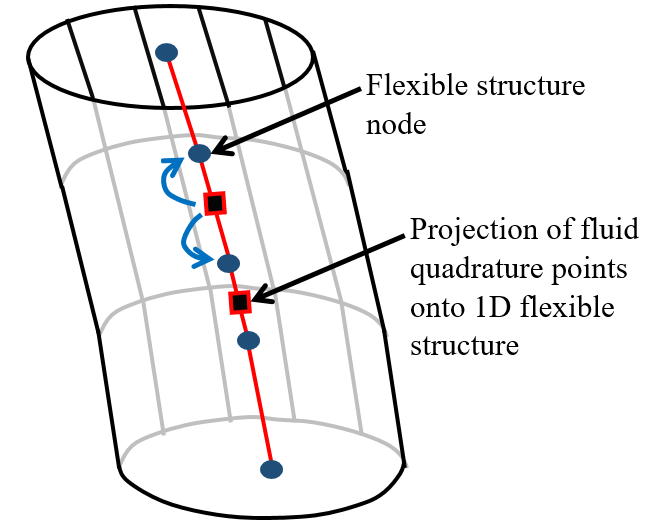}
		\caption{}
	\end{subfigure}
	\caption{Schematic of surface-to-line projection of traction: (a) quadrature projection of the fluid tractions from the fluid surface mesh to the corresponding structural line element and (b) evaluation of the fluid tractions at the structural nodes using the shape functions of the line element.}
	\label{fig:s-to-l}
\end{figure}
This section describes the transfer of the fluid tractions from the interface 3D fluid mesh elements onto the 1D flexible multibody line elements. A quadrature projection scheme is utilized to carry out this transfer in which the fluid tractions at the quadrature points of the fluid surface mesh are projected onto the targeted structural element as shown in Fig. \ref{fig:s-to-l}. Similar to the nodal projection method discussed before, in the quadrature projection scheme, the fluid surface mesh quadrature points are projected onto the structural line elements as shown in Fig. \ref{fig:s-to-l}a. Then these projected points are associated with elemental fluid tractions ($\boldsymbol{t}_q^\mathrm{f}$) evaluated at their corresponding quadrature points. Subsequently, the fluid traction acting on the $j^\mathrm{th}$ node of the 1D flexible multibody mesh is evaluated using the shape function ($\boldsymbol{\phi}_j^\mathrm{s}$) of the line element in the following manner:
	\begin{equation}
	\boldsymbol{t}^\mathrm{s}_{j}=\sum_{k=1}^{nf} \sum_{q=1}^{nq}  \int_{\gamma} \boldsymbol{\phi}_j^\mathrm{s}\left(\vec{\mathbb{P}}\left(\vec{x}^\mathrm{f}_q,t\right)\right) \boldsymbol{t}_q^\mathrm{f} \mathrm{d}\Gamma,
	\end{equation}
	where $\vec{x}^\mathrm{f}_q$ represents the position of the fluid mesh quadrature point, $\gamma$ is the 1D flexible multibody line element onto which quadrature point is projected, $nq$ is the number of quadrature points and $nf$ is number of nodes on the 1D flexible multibody mesh. From the summation property of the shape functions, the transfer of integrated traction load satisfies the conservation by construction.

\section{Partitioned Iterative Formulation for Surface-to-Line Coupling}
We now present NIFC scheme \cite{jaiman_CF16} which has been extended for the flexible multibody system considered in this study. We briefly describe the iterative force correction procedure for the multibody system. The linearized system of the coupled system which we get after the discretization and can be written in the matrix form as
\renewcommand\arraystretch{1}
\begin{align} \label{LS_Coupled}
	\begin{bmatrix}
		\boldsymbol{A}_{11} & 0 & 0 & \boldsymbol{A}_{14} \\
	    \boldsymbol{A}_{21} & \boldsymbol{A}_{22} & 0 & 0 \\
	    0 & \boldsymbol{A}_{32} & \boldsymbol{A}_{33} & 0 \\
	    0 & 0 & \boldsymbol{A}_{43} & \boldsymbol{A}_{44}
	\end{bmatrix} 
	\begin{Bmatrix}
		\Delta \boldsymbol{u}^\mathrm{s} \\
		\Delta \boldsymbol{u}^\mathrm{I} \\
		\Delta \boldsymbol{q}^\mathrm{f} \\
		\Delta \boldsymbol{f}^\mathrm{I}
	\end{Bmatrix}
	= \begin{Bmatrix} 
		\boldsymbol{R}_1 \\
		\boldsymbol{R}_2 \\
		\boldsymbol{R}_3 \\
		\boldsymbol{R}_4
	  \end{Bmatrix},
\end{align}
where the first equation corresponds to the structural multibody system, the second equation is the line-to-surface displacement mapping equation, the third equation deals with the ALE fluid and turbulence equation and the fourth equation is the surface-to-line traction continuity mapping. Here, $\Delta \boldsymbol{u}^\mathrm{s}$ is the solution increment for the structural displacement, $\Delta \boldsymbol{u}^\mathrm{I}$ is the displacement continuity increment at the fluid-structure interface, $\Delta \boldsymbol{q}^\mathrm{f}$ consists of the increments in $\bar{\boldsymbol{v}}^\mathrm{f}$, $\bar{p}$ and $\tilde{\nu}$ dealing with the Navier-Stokes and turbulence closure equations and $\Delta \boldsymbol{f}^\mathrm{I}$ is the increment in the traction continuity equation. The right-hand side vector $\boldsymbol{R}_i$ denotes the linear system residual for each equation $i$ of the coupled fluid, the interface displacement, the traction equilibrium and the multibody structural system.

To facilitate the staggered computation, we decouple the above equation to eliminate the off diagonal term $\boldsymbol{A}_{14}$. With the help of static condensation,
\begin{align} \label{NIFC_eq}
	\underbrace{(\boldsymbol{A}_{44} - \boldsymbol{A}_{43}\boldsymbol{A}_{33}^{-1}\boldsymbol{A}_{32}\boldsymbol{A}_{22}^{-1}\boldsymbol{A}_{21}\boldsymbol{A}_{11}^{-1}\boldsymbol{A}_{14} ) \Delta \boldsymbol{f}^\mathrm{I}}_{\tilde{\boldsymbol{A}}_{44}} = \nonumber \\
	 \underbrace{\boldsymbol{R}_4 - \boldsymbol{A}_{43}\boldsymbol{A}_{33}^{-1}( \boldsymbol{R}_3 - \boldsymbol{A}_{32}\boldsymbol{A}_{22}^{-1}( \boldsymbol{R}_2 - \boldsymbol{A}_{11}^{-1}\boldsymbol{A}_{21}\boldsymbol{R}_1) )}_{\tilde{\boldsymbol{R}}_4}
\end{align}
By substituting Eq.~(\ref{NIFC_eq}) in Eq.~(\ref{LS_Coupled}), we get
\renewcommand\arraystretch{1}
\begin{align}
	\begin{bmatrix}
		\boldsymbol{A}_{11} & 0 & 0 & 0 \\
	    \boldsymbol{A}_{21} & \boldsymbol{A}_{22} & 0 & 0 \\
	    0 & \boldsymbol{A}_{32} & \boldsymbol{A}_{33} & 0 \\
	    0 & 0 & 0 & \tilde{\boldsymbol{A}}_{44}
	\end{bmatrix} 
	\begin{Bmatrix}
		\Delta \boldsymbol{u}^\mathrm{s} \\
		\Delta \boldsymbol{u}^\mathrm{I} \\
		\Delta \boldsymbol{q}^\mathrm{f} \\
		\Delta \boldsymbol{f}^\mathrm{I}
	\end{Bmatrix}
	= \begin{Bmatrix} 
		\boldsymbol{R}_1 \\
		\boldsymbol{R}_2 \\
		\boldsymbol{R}_3 \\
		\tilde{\boldsymbol{R}}_4
	  \end{Bmatrix},
\end{align}
Since the Jacobian matrices in Eq.~(\ref{NIFC_eq}) are not directly available for staggered partitioned computation, an iterative procedure is formed to correct the tractions by a feedback process between the fluid and the structure. Therefore, the nonlinear iterative force correction is carried out by
\begin{align}
	\boldsymbol{f}^\mathrm{I}_{(\mathrm{k+1})} = \boldsymbol{f}^\mathrm{I}_{(\mathrm{k})} + \tilde{\boldsymbol{A}}_{44}^{-1}\tilde{\boldsymbol{R}}_{4(\mathrm{k})}
\end{align}
where $\mathrm{k}$ is the nonlinear iteration at a particular time step. The force correction vector is constructed by successive estimates without forming its inverse at each nonlinear iteration. This correction depends on an input-output relationship between the displacement from the structure and the force transfer from the fluid side and is evaluated by the generalization of Aitken's $\Delta^2$ extrapolation via dynamic weighting parameter to transform a fixed point iteration into a convergent and stable update of the forces at the fluid-structure interface \cite{jaiman_CF16}. This NIFC-based correction provides the necessary stability to the partitioned coupling for low structure-to-fluid mass ratio which is encountered typically for offshore systems.

\begin{figure}[H]
\centering
\begin{tikzpicture}[decoration={markings,mark=at position 0.5 with {\arrow[scale=2]{>}}},scale=1.0]
	\draw[-,black] (0,0) node[anchor=north,black]{$\Omega^\mathrm{f}$} to (0,11);
	\draw[-,black] (6,0) node[anchor=north,black]{$\Omega^\mathrm{s}_i$} to (6,11);
	\draw[dotted] (3,0) node[anchor=north,black]{$\Gamma^\mathrm{fs}_i$} to (3,11);

	\fill[black] (5.85,0.85) rectangle (6.15,1.15) ;
	\fill[black] (5.85,3.85) rectangle (6.15,4.15) ;
	\fill[black] (5.85,6.85) rectangle (6.15,7.15) ;
	\fill[black] (5.85,9.85) rectangle (6.15,10.15) ;

	\draw[black,fill=black] (0,1) circle (0.9ex);
	\draw[black,fill=black] (0,4) circle (0.9ex);
	\draw[black,fill=black] (0,7) circle (0.9ex);
	\draw[black,fill=black] (0,10) circle (0.9ex);
	
	\draw[black,postaction={decorate}] (0,1) to (6,1);
	\draw[black,postaction={decorate}] (4,0) to (6,1);
	\draw[black,postaction={decorate}] (0,10) to (2,11);

	\draw[black,postaction={decorate}] (6,1) to (6,4);
	\draw[black,postaction={decorate}] (6,4) to (0,1);
	\draw[black,postaction={decorate}] (0,1) to (0,4);
	\draw[black,postaction={decorate}] (0,4) to (6,4);

	\draw[black,postaction={decorate}] (6,4) to (6,7);
	\draw[black,postaction={decorate}] (6,7) to (0,4);
	\draw[black,postaction={decorate}] (0,4) to (0,7);
	\draw[black,postaction={decorate}] (0,7) to (6,7);

	\draw[black,postaction={decorate}] (6,7) to (6,10);
	\draw[black,postaction={decorate}] (6,10) to (0,7);
	\draw[black,postaction={decorate}] (0,7) to (0,10);
	\draw[black,postaction={decorate}] (0,10) to (6,10);

	\draw[black,thick] (-0.25,7.6) to (0.25,7.8);
	\draw[black,thick] (-0.25,7.75) to (0.25,7.95);
	\draw[white] (0,7.72) to (0,7.83) ;

	\draw[black,thick] (5.75,7.6) to (6.25,7.8);
	\draw[black,thick] (5.75,7.75) to (6.25,7.95);
	\draw[white] (6,7.72) to (6,7.83) ;
	\draw[black,dashed] (6,1) to (9,1);
    \draw[black,dashed] (6,10) to (9,10);

	\draw[black] (6.2,0.5) node[anchor=west,black]{$\boldsymbol{u}^\mathrm{s,-}_\mathrm{n}(\boldsymbol{X}^\mathrm{s},t_\mathrm{n}^{-})$};
		\draw[black] (6.2,1.5) node[anchor=west,black]{$\boldsymbol{u}^\mathrm{s,+}_\mathrm{n}(\boldsymbol{X}^\mathrm{s},t_\mathrm{n}^{+})$};
	\draw[black] (7.3,10.5) node[anchor=west,black]{$\boldsymbol{u}^\mathrm{s,+}_\mathrm{n+1}(\boldsymbol{X}^\mathrm{s},t_\mathrm{n+1}^{+})$};
		\draw[black] (7.3,9.5) node[anchor=west,black]{$\boldsymbol{u}^\mathrm{s,-}_\mathrm{n+1}(\boldsymbol{X}^\mathrm{s},t_\mathrm{n+1}^{-})$};
\draw [black, xshift=4cm] plot [smooth, tension=0.75] coordinates { (4.6,1) (4.4,3.5) (3.2,7.5) (3,10)};
\draw [black, xshift=4cm] plot [smooth, tension=0.75] coordinates { (3.3,10) (3.2,10.5) (2.9,11)};
\draw [black, xshift=4cm] plot [smooth, tension=0.75] coordinates { (4.7,0) (4.4,0.5) (4.3,1)};
	\draw[black] (-0.2,1) node[anchor=east,black]{$\bar{\boldsymbol{v}}^\mathrm{f}(\boldsymbol{x}^\mathrm{f},t^\mathrm{n})$,};
	\draw[black] (-0.2,0.5) node[anchor=east,black]{$\bar{p}(\boldsymbol{x}^\mathrm{f},t^\mathrm{n})$,};
 	\draw[black] (-0.2,-0.1) node[anchor=east,black]{$\tilde{\nu}^\mathrm{f}(\boldsymbol{x}^\mathrm{f},t^\mathrm{n})$};

	\draw[black] (-0.2,11.1) node[anchor=east,black]{$\bar{\boldsymbol{v}}^\mathrm{f}(\boldsymbol{x}^\mathrm{f},t^\mathrm{n+1})$,};
 	\draw[black] (-0.2,10.6) node[anchor=east,black]{$\bar{p}(\boldsymbol{x}^\mathrm{f},t^\mathrm{n+1})$,};
 	\draw[black] (-0.2,10) node[anchor=east,black]{$\tilde{\nu}^\mathrm{f}(\boldsymbol{x}^\mathrm{f},t^\mathrm{n+1})$};

	\draw (6,4) -- (0,1) node [midway, above, sloped] {$\bar{\boldsymbol{v}}^\mathrm{f} = \boldsymbol{v}^\mathrm{s}$};
	\draw (2.3,4.3) node[anchor=west]{$\boldsymbol{f}^\mathrm{s}_{\mathrm{k+1}} = \boldsymbol{f}^\mathrm{s}_{\mathrm{k}} + \delta\boldsymbol{f}^\mathrm{s}_\mathrm{k}$};
	\draw (6,2.5) node[anchor=west,black]{(1)};
	\draw (3.4,2.5) node[anchor=west,black]{(2)};
	\draw (0,2.5) node[anchor=east,black]{(3)};
	\draw (3.9,3.7) node[anchor=east,black]{(4)};

	\draw[black] (8.8,8.5) node{(1)};
	\draw (9,8.5) node[anchor=west,black]{Solve multibody};
	\draw[black] (8.8,7.5) node{};
	\draw (9,8) node[anchor=west,black]{system with constraints};
	\draw[black] (8.8,7) node{(2)};
	\draw (9,7) node[anchor=west,black]{Map structural disp- };
	\draw[black] (8.8,6.5) node{};
	\draw (9,6.5) node[anchor=west,black]{lacement and velocity};
	\draw[black] (8.8,6) node{};
	\draw (9,6) node[anchor=west,black]{on the 3D fluid mesh};
	\draw[black] (8.8,5) node{(3)};
	\draw (9,5) node[anchor=west,black]{Solve ALE fluid and};
	\draw[black] (8.8,4.5) node{};
	\draw (9,4.5) node[anchor=west,black]{turbulence equations};
	\draw[black] (8.8,3.5) node{(4)};
	\draw (9,3.5) node[anchor=west,black]{Compute the forces and};
	\draw[black] (8.8,3) node{};
	\draw (9,3) node[anchor=west,black]{map them on the line};
	\draw[black] (8.8,2.5) node{};
	\draw (9,2.5) node[anchor=west,black]{elements};
	
	\draw (1.5,2.5) node[anchor=east]{$\mathrm{k}=1$};
	\draw (1.5,5.5) node[anchor=east]{$\mathrm{k}=2$};
	\draw (2,8.5) node[anchor=east]{$\mathrm{k}=\mathrm{n_{iter}}$};

	\draw (6.7,2.8) node[rotate=90, anchor=west,black]{Flexible-multibody system};
	\draw (-1.0,7.0) node[rotate=90, anchor=east,black]{ALE fluid \& turbulence};
\end{tikzpicture}
\caption{A schematic of predictor-corrector procedure of proposed NIFC scheme that couples fluid solver and multibody solver. For the flow solver an unconditionally stable and second-order accurate generalized-$\alpha$ method is utilized for the time integration, whereas, for the multibody solver a time discontinuous Galerkin scheme based on energy decay inequality is used.}
\label{fig:solver}
\end{figure}
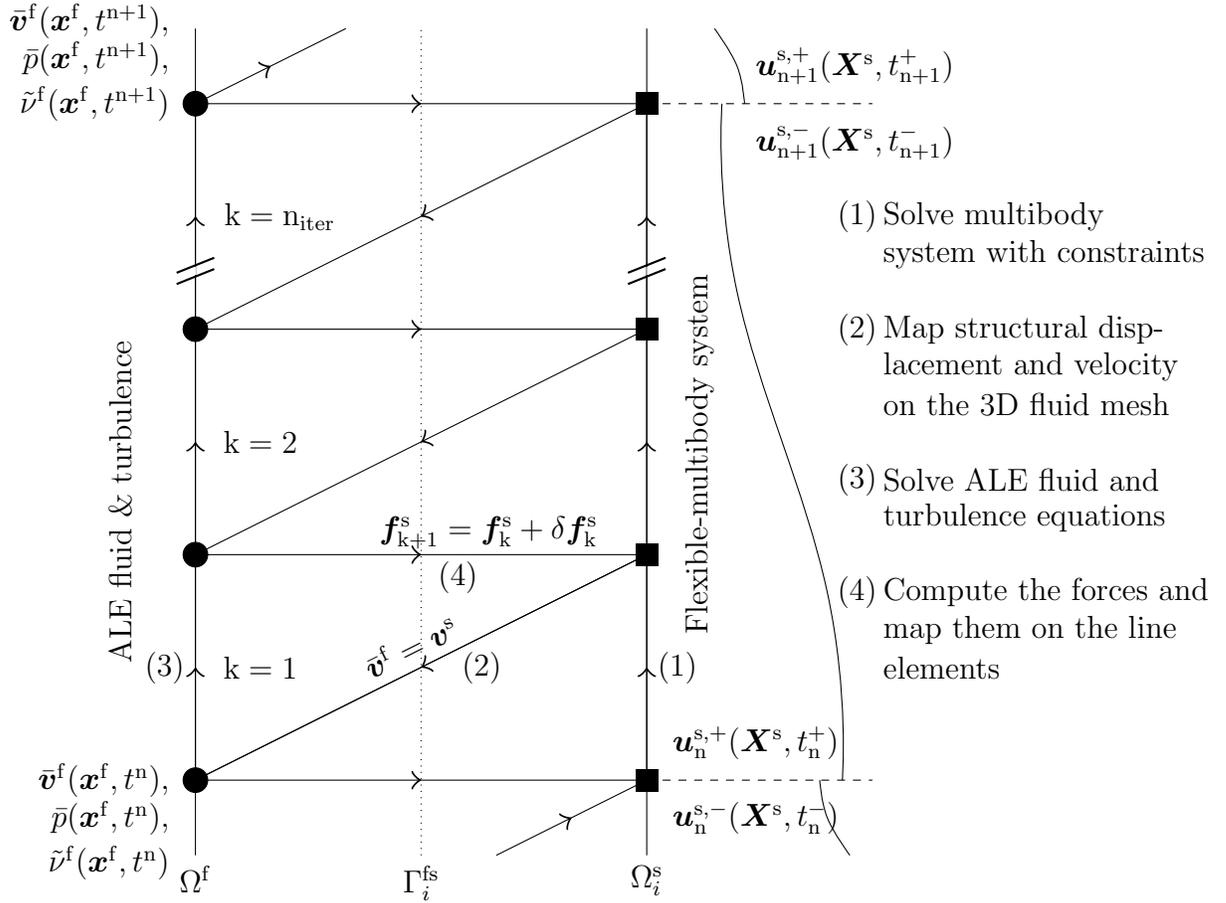

A schematic of the coupling procedure is shown in Fig. \ref{fig:solver}. The structural update provides the predictor displacement for the FMI solver and the fluid solver acts as a corrector step to construct the forces at the fluid-structure interface. Consider the structural displacements $\boldsymbol{u}^\mathrm{s,+}_\mathrm{n}$ and $\boldsymbol{u}^\mathrm{s,-}_\mathrm{n}$ at time $t^{+}_\mathrm{n}$ and $t^{-}_\mathrm{n}$ respectively due to the fluid forces at time $t^\mathrm{n}$. In the first step of the algorithm, we first solve for the structural displacement for each multibody using the given computed fluid forces at time $t^\mathrm{n}$ by employing the time discontinuous Galerkin approximation. For the present study, the time step size for both the structural and fluid solvers is chosen to be identical, i.e., $\Delta t= \Delta t^\mathrm{s}= \Delta t^\mathrm{f}$. However, they can be different leading to a partitioned staggered type of coupling. In the second step of the nonlinear iteration $\mathrm{k}$, the predicted structural displacement is then transferred to the fluid solver by satisfying the ALE geometric compatibility and the velocity continuity at the interface $\Gamma^\mathrm{fs}$ using the line-to-surface coupling. The flow equations under the ALE framework along with the turbulence closure equations are solved as the third step of the iteration $\mathrm{k}$. In the final step, the evaluated fluid forces are then iteratively corrected using the nonlinear iterative force correction (NIFC) technique, after which, the updated forces are transferred to the structure side via the surface-to-line coupling. The FMI solver then moves to the next nonlinear iteration. When the solver has achieved the convergence criteria, all the variables are updated for the next time step $t^\mathrm{n+1}$.

The resulting coupled algebraic system of flow equations obtained from the finite element discretization are solved via the Generalized Minimal RESidual (GMRES) algorithm \cite{saad1986} which relies on the Krylov subspace iteration and the modified Gram-Schmidt orthogonalization. In the present framework, a Krylov space of 30 orthonormal vectors is utilized to solve the coupled ALE fluid flow (pressure and velocity) and the turbulence matrix system along with the diagonal preconditioners. At each time step, Newton-Raphson type iterations are used to minimize the linearization error. Similarly, the system of algebraic equation obtained by discretization of flexible multibody equations in the co-rotational framework are solved using the classical skyline solver which is based 
on the factorization of the system matrix \cite{bauchau2007dymore}. 
In the coupled flexible-mulitbody solver, the flow-turbulence computations 
are performed via message passing interface (MPI) and domain decomposition strategy \cite{mpi,smith_mpi} 
on distributed memory clusters, while the finite element computation of multibody 
structural solver is done in a serial manner on a single compute node.

\section{Validation of Vortex-Induced Vibration of Flexible Cylinder}\label{Val}
\begin{figure}[!htbp]
	\centering
	\begin{subfigure}[b]{0.3\textwidth}
		\begin{tikzpicture}[every node/.style={scale=0.8},scale=0.8]
	\draw [thick] (-0.1,0) to (-0.1,11);
	\draw [thick] (0.1,0) to (0.1,11);
	\draw [thick] (-0.1,0) to (0.1,0);
	\draw [thick] (-0.1,11) to (0.1,11);

	\draw [thick] (-0.2,-0.5) to (0.2,-0.5);
	\draw [thick] (-0.2,-0.5) to (0,0) ;
	\draw [thick] (0.2,-0.5) to (0,0) ;
	\draw [thick] (-0.2,11.5) to (0.2,11.5);
	\draw [thick] (-0.2,11.5) to (0,11) ;
	\draw [thick] (0.2,11.5) to (0,11) ;
	
	\draw [thick,->] (0,11.5) -- (0,12.2);
	\draw (0,12.2) node[right]{$P$};

	\draw [thick] (-2,0) to (-2,11);
	\draw [->,thick] (-2,0) to (-1.5,0);
	\draw [->,thick] (-2,1) to (-1.5,1);
	\draw [->,thick] (-2,2) to (-1.5,2);
	\draw [->,thick] (-2,3) to (-1.5,3);
	\draw [->,thick] (-2,4) to (-1.5,4);
	\draw [->,thick] (-2,5) to (-1.5,5);
	\draw [->,thick] (-2,6) to (-1.5,6);
	\draw [->,thick] (-2,7) to (-1.5,7);
	\draw [->,thick] (-2,8) to (-1.5,8);
	\draw [->,thick] (-2,9) to (-1.5,9);
	\draw [->,thick] (-2,10) to (-1.5,10);
	\draw [->,thick] (-2,11) to (-1.5,11);
	\draw (-2,5) node[left]{$U$};

	\draw [->,thick] (-0.5,5.5) to (-0.1,5.5);
	\draw [->,thick] (0.8,5.5) to (0.1,5.5);
	\draw (0.6,5.5) node[above]{$D$};


	\draw [->,thick] (-1.1,0) to (-0.2,0);
	\draw (-1,.2) node[right]{$X$};
	\draw [->,thick] (-1.1,0.1) to (-1.1,1);
	\draw (-1,1) node[above]{$Z$};
	\draw [thick] (-1.1,0) circle (0.1cm);
	\draw (-1,-0.1) node[below]{$Y$};
	\draw [thick] (-1.03,0.0707) to (-1.17,-0.0707);	
	\draw [thick] (-1.03,-0.0707) to (-1.17,0.0707);
\end{tikzpicture}
	\caption{}
	\label{schematic_setup}
	\end{subfigure}%
	\begin{subfigure}[b]{0.5\textwidth}
\begin{tikzpicture}[very thick,decoration={markings,mark=at position 0.5 with {\arrow{>}}},every node/.style={scale=0.6},scale=0.6]
	\draw (0,0) node[left]{} -- (0,15) node[right]{} -- (8,15) node[above]{} -- (8,0) node[above]{} -- cycle;
	\draw[black,dotted] (1,1) to (1,16);
	\draw[black] (1,16) to (9,16);
	\draw[black] (9,16) to (9,1);
	\draw[black,dotted] (9,1) to (1,1);

	\draw[black] (0,15) to (1,16);
	\draw[black] (8,15) to (9,16);
	\draw[black] (8,0) to (9,1);
	\draw[black,dotted] (0,0) to (1,1);

	\draw[dotted] (3,15.5) to (3,0.5);
	\draw[dotted] (3.2,15.5) to (3.2,0.5);
	\draw (3.1,15.5) circle (0.1cm) ;
	\draw[dotted] (3.1,0.5) circle (0.1cm);

	\draw[->](-2.5,8.5) to (-0.5,8.5);
	\draw (-1.5,9.5) node[anchor=north]{Inlet};

	\draw (9.2,1) to (9.6,1);
	\draw (9.2,16) to (9.6,16);
	\draw[<->] (9.4,1) to (9.4,16);
	\draw (10.8,9.2) node[anchor=north]{Outlet};
	\draw (9.4,12) node[anchor=west]{$481.5D$};

	\draw (5,17.2) node[anchor=north]{Top (Slip boundary)};

	\draw (5.5,2.5) node[anchor=north]{Bottom};
	\draw (5.3,2) node[anchor=north]{(Slip boundary)};

	\draw[<-] (3.2,8.5) to (4,7.5);
	\draw (4,7.5) to (4.25,7.5);
	\draw (4.25,7.5) node[anchor=west]{No-slip};

	\draw (0,-0.2) to (0,-0.6);
	\draw (3.1,-0.2) to (3.1,-0.6);
	\draw (8,-0.2) to (8,-0.6);
	\draw[<->] (0,-0.4) to (3.1,-0.4);
	\draw (1.5,-0.4) node[anchor=north]{$10D$};
	\draw[<->] (3.1,-0.4) to (8,-0.4);
	\draw (5.5,-0.4) node[anchor=north]{$25D$};
	\draw (8.2,0) to (8.6,0);
	\draw[<->] (8.4,0) to (9.4,1);
	\draw (8.9,0.4) node[anchor=west]{$20D$};

	\draw[->,red] (-2,15) to (-1,15);
	\draw (-1,15) node[anchor=west,red]{X};
	\draw[->,red] (-2,15) to (-2,16);
	\draw (-2,16) node[anchor=south,red]{Z};
	\draw[->,red] (-2,15) to (-1.29,15.71);
	\draw (-1.4,15.71) node[anchor=west,red]{Y};
\end{tikzpicture}
	\caption{}
	\label{schematic_comp}
	\end{subfigure}%
	\caption{A long flexible riser model in a uniform current flow along the $Z$-axis: 
        (a) pinned-pinned tensioned riser with uniform flow, (b) schematic illustration of the computational setup and boundary conditions.}
	\label{schematic}
\end{figure}
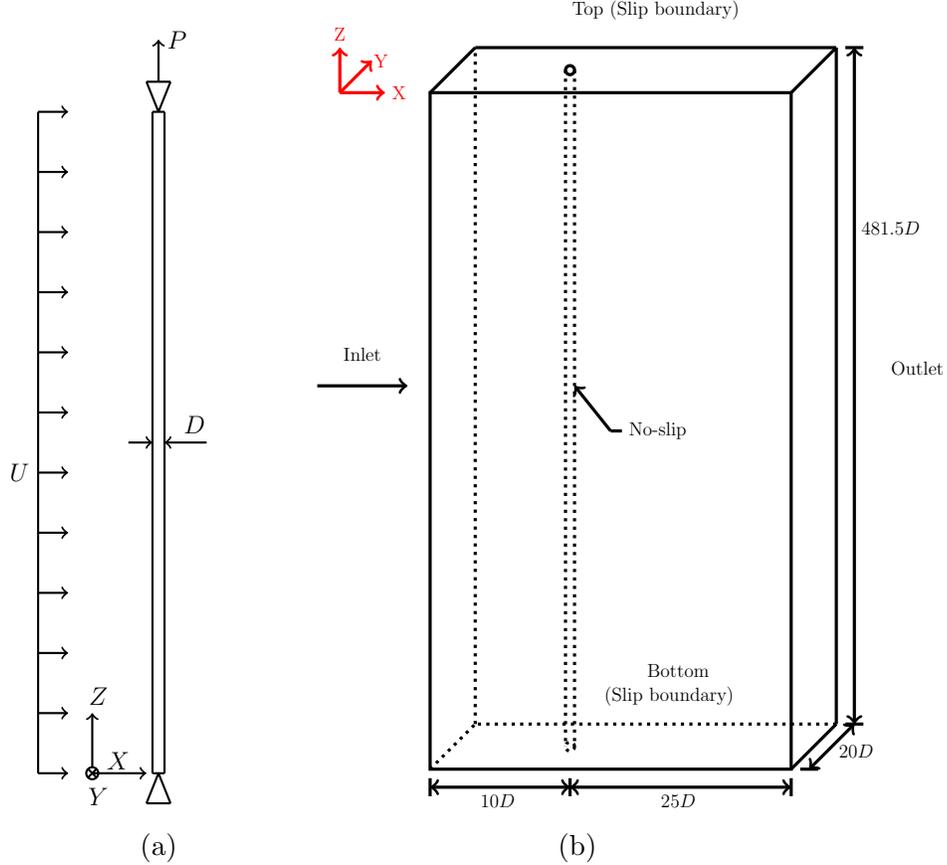

In this section, the validation of the proposed coupling between the flow solver and the multibody solver under the NIFC framework is performed for vortex-induced vibration of a flexible cylinder (offshore riser) with the same setup as presented in \cite{Vaibhav_CAF}. A schematic description of the riser is shown in Fig. \ref{schematic_setup} and the corresponding computational domain for the simulation is presented in Fig. \ref{schematic_comp}. In the experiments \cite{ExxonMobil}, a pre-tensioned flexible riser with pinned-pinned boundary condition was subjected to a uniform current of $0.2$ m/s  and its response characteristics were measured.

The riser spans $481.5D$ in the $Z$-direction, where $D$ is the diameter of the riser. The inlet and outlet of the computational domain is placed at a distance $10D$ and $25D$ from the center of the riser respectively. The side walls are placed equidistant from the center of the riser with a distance of $10D$ on each side with a blockage of $5\%$. No-slip boundary condition with $\bar{\nu} = 0$ is imposed on the riser wall, whereas slip conditions are satisfied at the side-walls and planes perpendicular to the axis of the riser, i.e., top and bottom. The free-stream velocity at the inlet boundary is along the $X$-axis with $\bar{\nu} = 0$ which corresponds to no incoming turbulence.  The non-dimensional parameters used for the simulation are presented in Table \ref{riser:tab}.
\begin{table}[]
	\centering
	\caption{Non-dimensional parameters used in riser VIV simulation}
		\begin{tabular}{@{}lc@{}}
			\hline
			Parameters &value\\ \hline
			\hline
	$K_B \left (=\frac{EI}{\rho^\mathrm{f} U^2D^4} \right) $ & $ 2.1158 \times 10^7$ \\
	$K_T \left(=\frac{P}{\rho^\mathrm{f} U^2D^2} \right) $ & $5.10625 \times 10^4$  \\
	$Re \left( =\frac{\rho^\mathrm{f} UD}{\mu^\mathrm{f}} \right) $ & $ 4000$  \\
	$\rho^* \left( =\frac{\rho^\mathrm{s}}{\rho^\mathrm{f}} \right)$ & $ 2.23$ \\
			\hline
		\end{tabular}\label{riser:tab}
	\end{table}
The variables in Table \ref{riser:tab} symbolized by $Re$ and $\rho^*$ are the Reynolds number and density ratio respectively with $\rho^\mathrm{s}$ being the density of the riser.

The fluid domain is discretized into approximately 3.5 million nodes with an unstructured finite element mesh of 8-node brick elements. The mesh characteristics are chosen based on the convergence study provided in \cite{Vaibhav_CAF} and it is similar to $M1$ mesh used therein. The structural domain, i.e., the riser is discretized with $200$ nonlinear beam elements. Boundary layer and wake regions are resolved sufficiently in the $X$-$Y$ plane and the mesh is kept the same and the number of 
spanwise layer is taken as $200$ for the present case. The boundary layer thickness of the riser is selected as $0.25D$ with the stretching ratio, $\Delta y_{j+1}/\Delta y_j$ of $1.15$ and the number of divisions in the wall-normal direction is chosen in a manner such that $y^+ < 1$. The non-dimensional time step size ($\Delta t U/D$) is selected as $0.1$ for the present problem.

\begin{figure}[!htbp]
	\centering
	\begin{subfigure}{.5\textwidth}
		\centering
		\includegraphics[trim={11cm 0cm 11cm 1cm},clip,width=7cm]{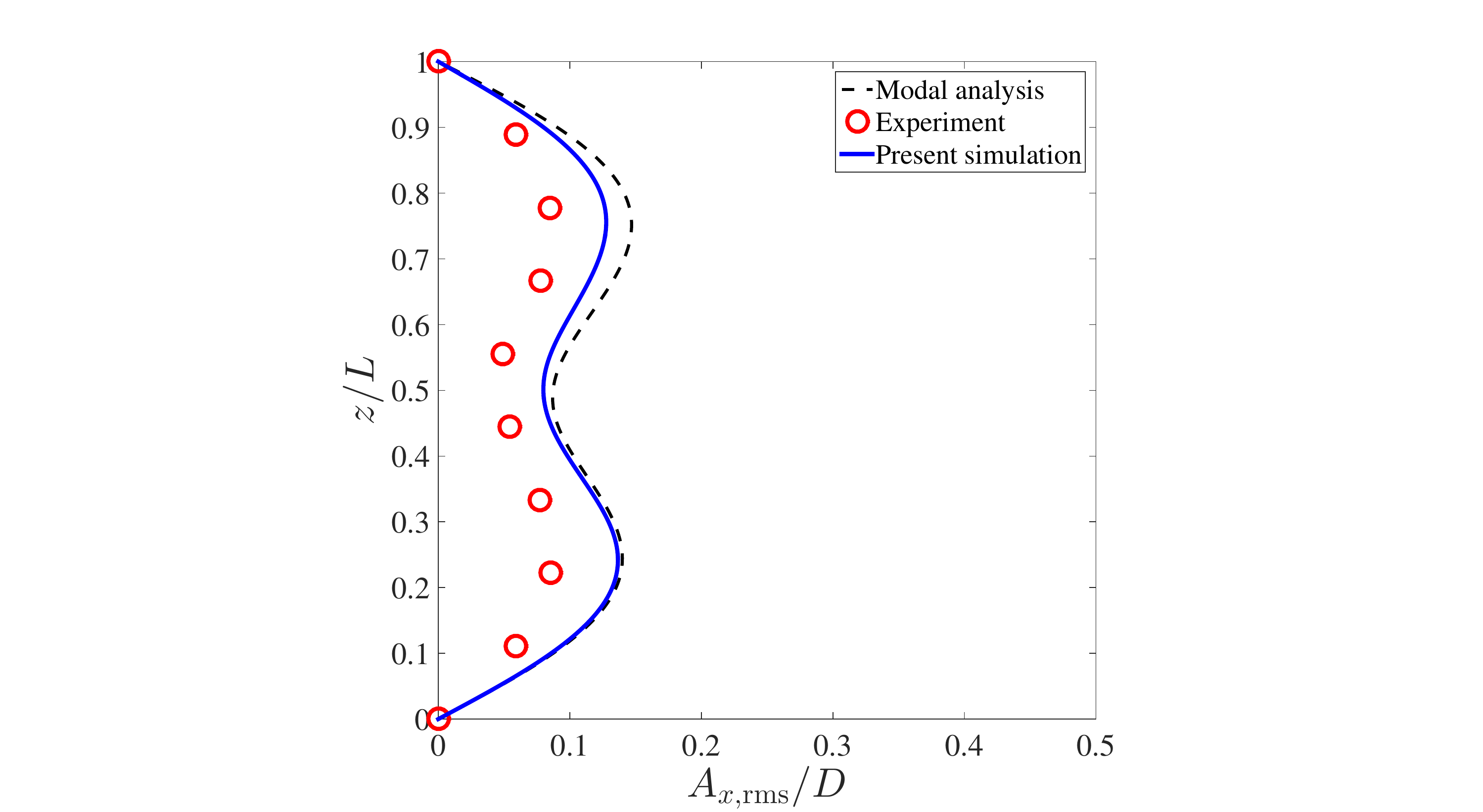}
		\caption{}
		\label{riser:rms_y}
	\end{subfigure}%
	\begin{subfigure}{.5\textwidth}
		\centering
		\includegraphics[trim={11cm 0cm 11cm 1cm},clip,width=7cm]{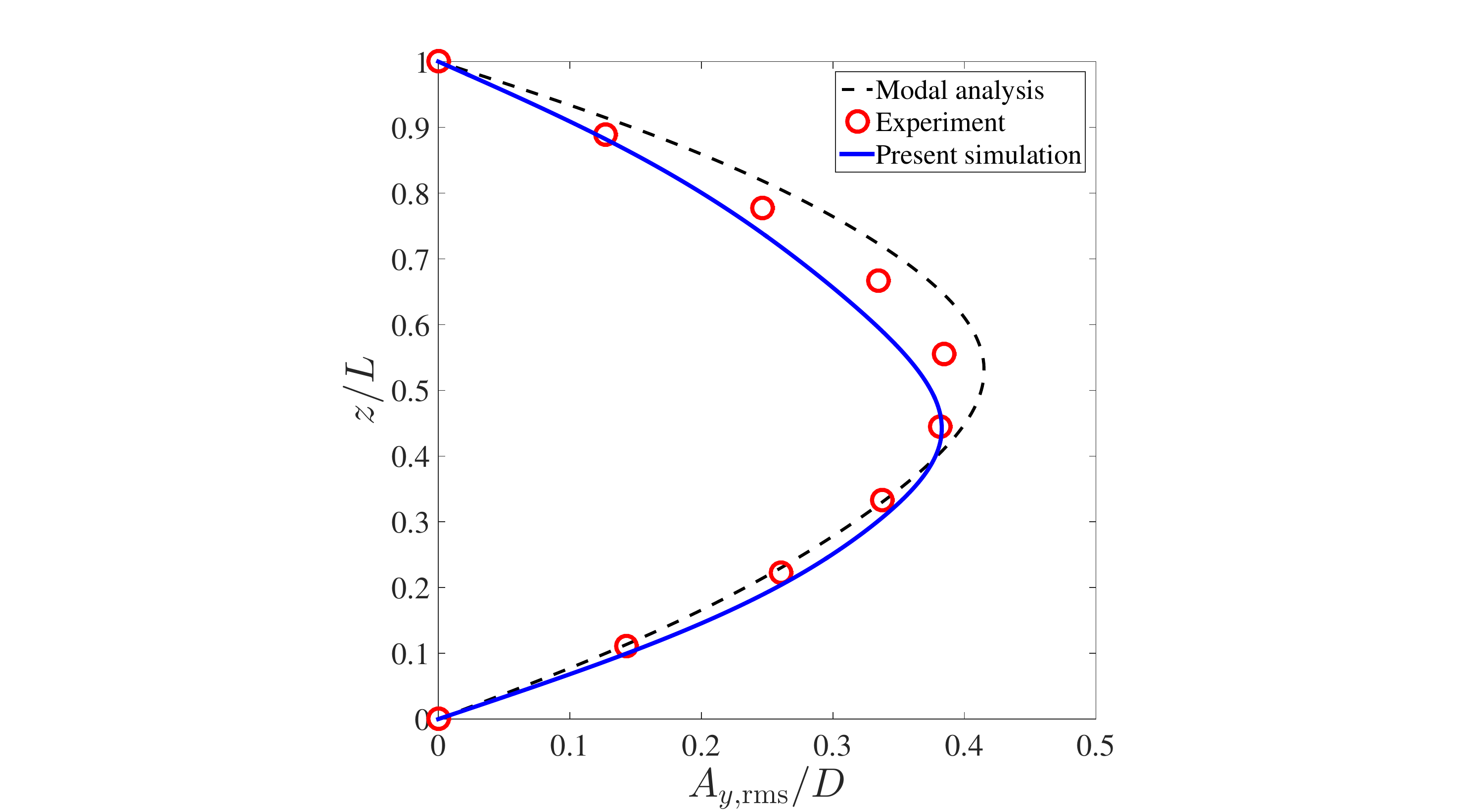}
		\caption{}
		\label{riser:rms_x}
	\end{subfigure}
	\caption{Rms-amplitudes of displacements for uniform current flow past a flexible riser modeled using nonlinear beam at ($Re$;$m^*$) = (4000; 2.23): (a) in-line and (b) cross-flow directions. \mbox{(\hdashrule[0.5ex]{0.8cm}{0.4mm}{1.5mm}) Modal analysis \cite{Vaibhav_CAF}}, \mbox{(\redcolor{\boldsymbol{$\circ$}}) Experiment \cite{ExxonMobil}} and \mbox{(\bluecolor{\hdashrule[0.5ex]{0.5cm}{0.4mm}{0.001mm}}) Present simulation}.  The riser is vibrating in the fundamental mode in the CF and the second mode for the IL directions.}
	\label{riser:rms}
\end{figure}

\begin{figure}[!htbp]
	\centering
	\begin{subfigure}[b]{\textwidth}
		\includegraphics[width=\linewidth]{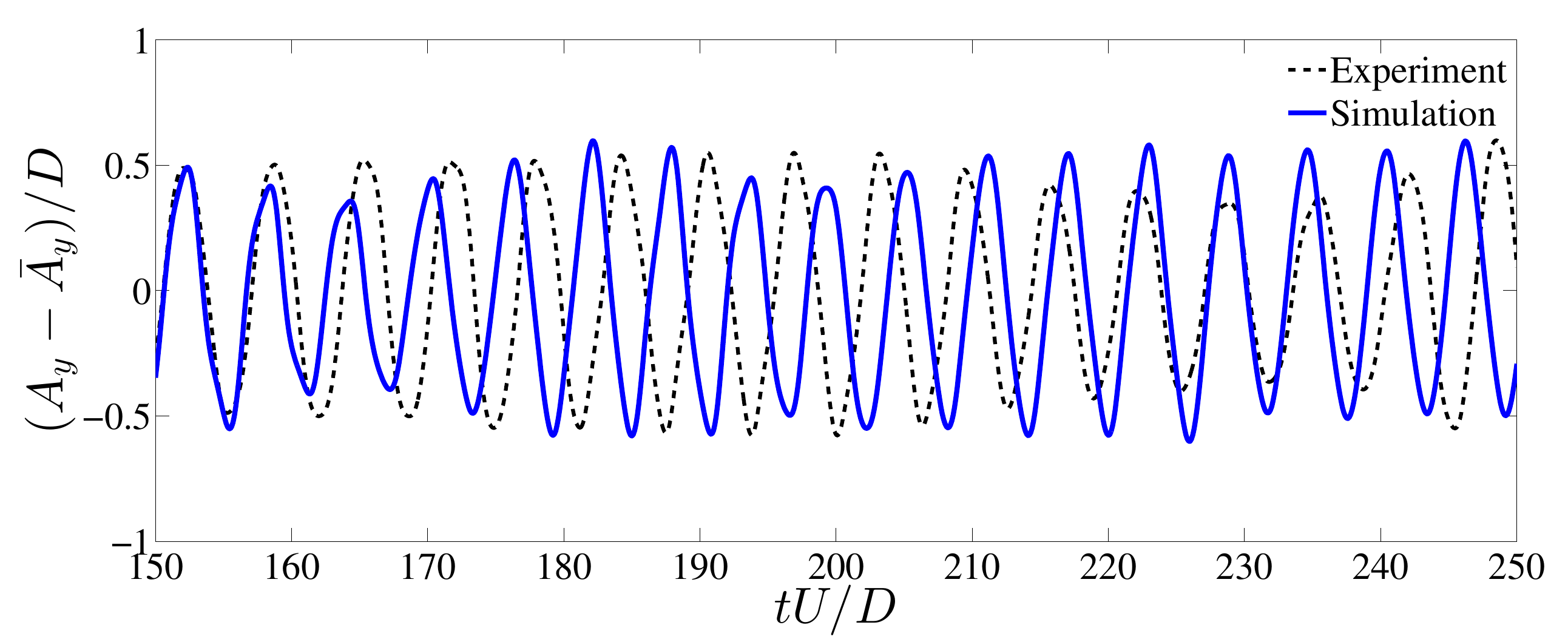}
		\caption{}
	\end{subfigure}%
	\caption{Uniform current flow past a flexible riser at $Re=4000$: comparison of the cross-flow response at $z/L=0.55$ with that of the experiment data for identical parameters.}
	\label{riser:cross_time}
\end{figure}

The riser response root mean square (rms) amplitudes of the displacements along the riser in both in-line (IL) and cross-flow (CF) directions are shown in Fig. \ref{riser:rms}. It can be observed that the rms-amplitudes of displacement in the cross-flow direction agree quite well with the experiment (the difference is less than $1\%$). However, the percentage of error in the in-line direction is high compared to the cross-flow direction $\sim$ $10\%$. This difference may be attributed to the geometric imperfections of the riser surface in the experiments and subsequent complexities in the flow separation. Moreover, the numerical prediction and measurement of the in-line response are very sensitive to the precise lock-in range and the boundary layer characteristics around the vibrating flexible riser. From a practical viewpoint, the in-line response is several factors smaller than the cross-flow amplitude, hence a good estimate of the cross-flow response is generally sufficient for the riser design study. A comparison of the riser response with the study employing modal analysis carried out in \cite{Vaibhav_CAF} is also made in Fig. \ref{riser:rms}. In Fig. \ref{riser:cross_time}, the time history of the cross-flow displacement along the riser at position $z/L = 0.55$ obtained from both the present simulation and the experiment is compared. It can be observed that the cross-flow response amplitude is in good agreement with that of the experimental measurements.

\begin{figure}[!htbp]
	\centering
	\begin{subfigure}{.5\textwidth}
		\centering
		\includegraphics[width=\linewidth]{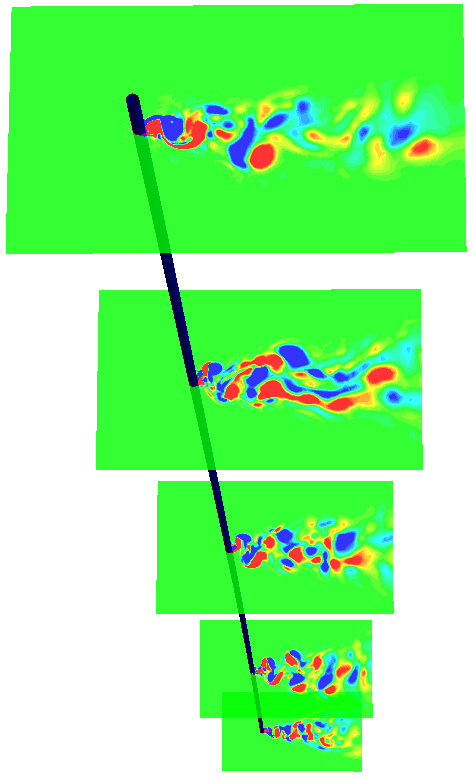}
		\caption{}
		\label{riser:vortex_plane}
	\end{subfigure}%
	\begin{subfigure}{.4\textwidth}
		\centering
		\includegraphics[width=\linewidth]{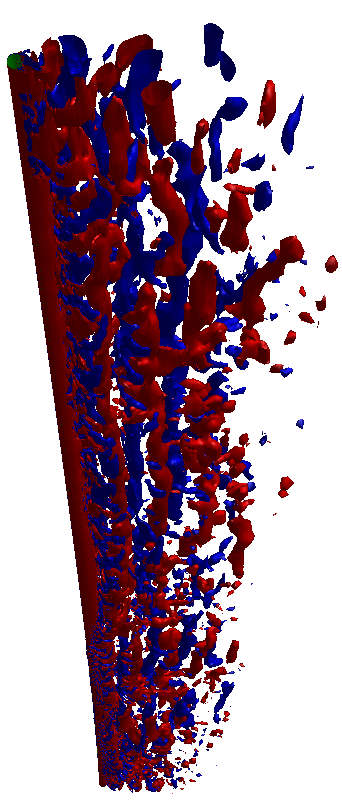}
		\caption{}
		\label{riser:vortex_3D}
	\end{subfigure}
	\caption{Vortex patterns (at $tU/D=250$) formed due to the flow induced vibration of flexible riser modeled using nonlinear beam: (a) $Z$-vorticity contours in various spanwise sections, red and blue color indicate the positive and negative vorticity respectively, and  (b) instantaneous iso-surfaces of vortical structures.}
	\label{riser:vortex}
\end{figure}

The flow pattern along the riser is visualized by plotting the two-dimensional vortex structures as well as the iso-surfaces of  the vortical structures  using a vortex-identification based on
 $Q$-criterion in Fig. \ref{riser:vortex}. The vortex patterns formed due to the flow inducted vibration of the flexible riser is presented for $tU/D=250$. The pattern is very complex, however, in general 2S mode of vortex shedding is observed in most of the locations. The locations where the amplitude of vibration is large, a wider 2S with two rows configuration is observed. A close observation of the Fig. \ref{riser:vortex_3D} shows that the iso-surfaces of the 3D vortices are divided into upper half and a lower half along the spanwise direction. It is observed that the vortices form a tube-like shape around the anti nodes of the vibration ($z/L = 0.25$ and $0.75$); whereas smaller vortices are formed at the node of the in-line vibration ($z/L = 0.5$) and near the supports. Similar observations were reported in \cite{Wang,Vaibhav_CAF} and hence it can be inferred that the proposed NIFC framework for the line-to-surface  and vice-versa coupling for the flexible bodies of a multibody system performs quite well in capturing the flow physics as well as the response characteristics.

\section{Application to Coupled Floater-Mooring-Riser System}\label{App}
\begin{figure}
	\centering
	\begin{subfigure}{0.8\textwidth}
		\centering
		\includegraphics[width=0.98\columnwidth]{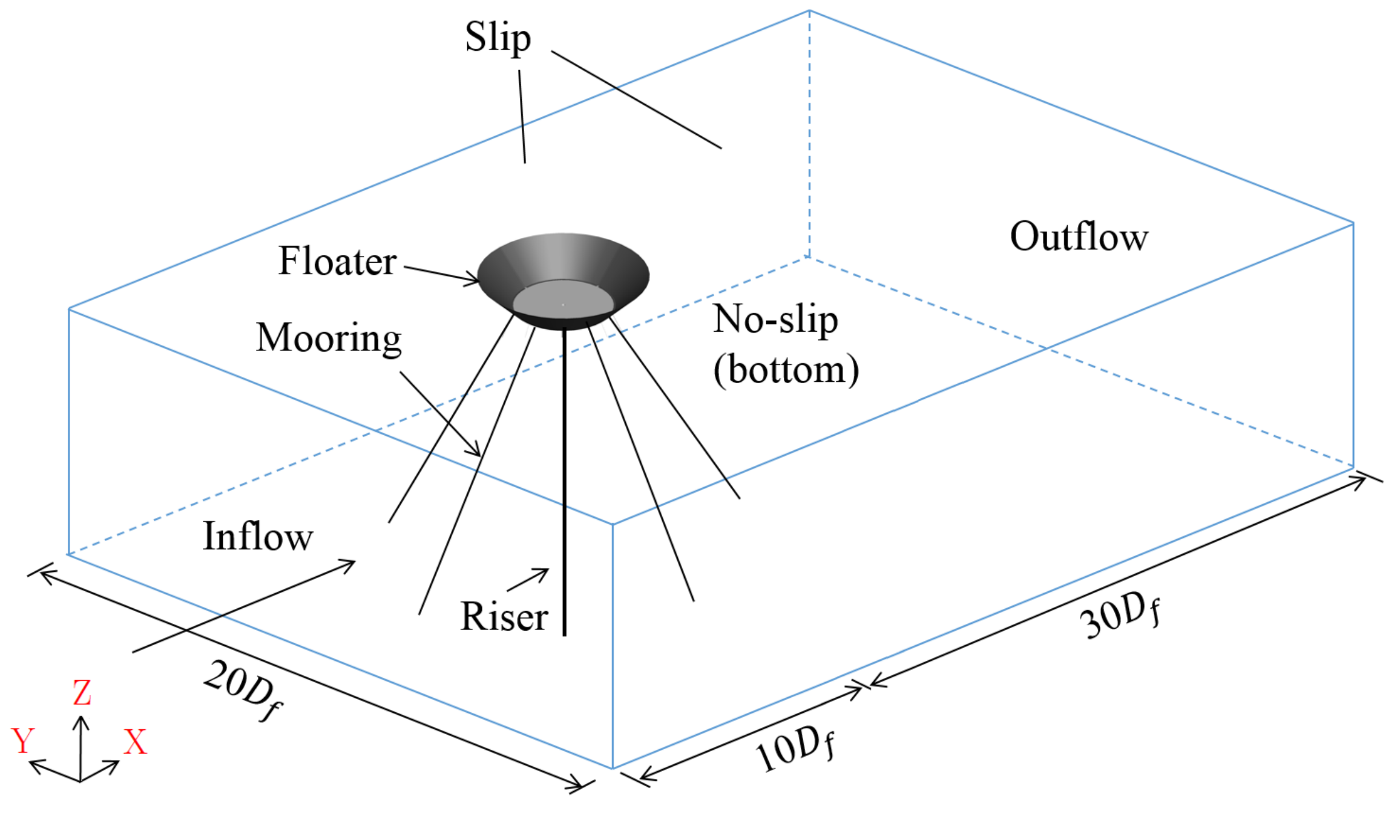}
	\end{subfigure}
	\caption{Schematic illustration of the computational domain of the floater-mooring-riser system subjected to uniform current flow in the $X$ direction.}
	\label{fig:panel}
	\end{figure}

In this section, a practical demonstration of the proposed numerical framework for an ocean engineering application of the floater-riser-mooring system is presented. Floaters are typically large mass body which are generally modeled as rigid bodies and can undergo vortex-induced motion in high ocean currents. Mooring lines are used for station-keeping of the floating system with one end connected to the floater through joints/constraints and the other end is anchored to the ocean floor. Owing to negligible flexural rigidity, the mooring lines are modeled using cable elements. Risers are long elastic pipes that connect an offshore production system to a drilling rig and/or sub-sea system and typically modeled as beam element. In Fig. \ref{fig:panel}, a schematic illustration of the floater-mooring-riser system in a uniform flow is shown. Design and analysis of the coupled system can be very challenging due to the complex floater motion exposed to the environmental forces and its connection with the moorings and riser system, which can have their own local vortex-induced forces.

For the present demonstration, the floater-mooring-riser system consists of a rigid funnel-shaped floater, four taut mooring lines and a long flexible riser. The floater corresponds to a typical Arctic hull with a downward ice-breaking slope having a diameter at the water-plane $D_f$ which is 50 times the diameter of the riser ($D_r$),  i.e., $D_f=50D_r$.  A draft of 20$D_f$ is considered for the floater. Mooring lines are equally placed in a radial direction with $90\deg$ azimuth angles facing the current direction. Moorings are $538.33D_r$ long having a diameter of $0.1D_r$. These are clamped at the bottom edge of the floater and extended till seabed which is at a distance $481.5D_r$ vertically from the bottom of the floater. A riser is vertically connected at the center of the lower panel of floater (Fig. \ref{fig:panel}) spanning $481.5D_r$ in $Z$-direction. All the moorings and riser are clamped to the floater such that the motion of the floater and top nodes of the moorings/risers are identical. The pinned boundary condition is used at seabed for all the moorings and the riser. All the  dimensions are chosen such that the flexible bodies do not encounter contact with each other. 
	\begin{figure}	
		\centering
		\includegraphics[width=0.98\columnwidth,trim=0 110mm 0 22mm, clip]{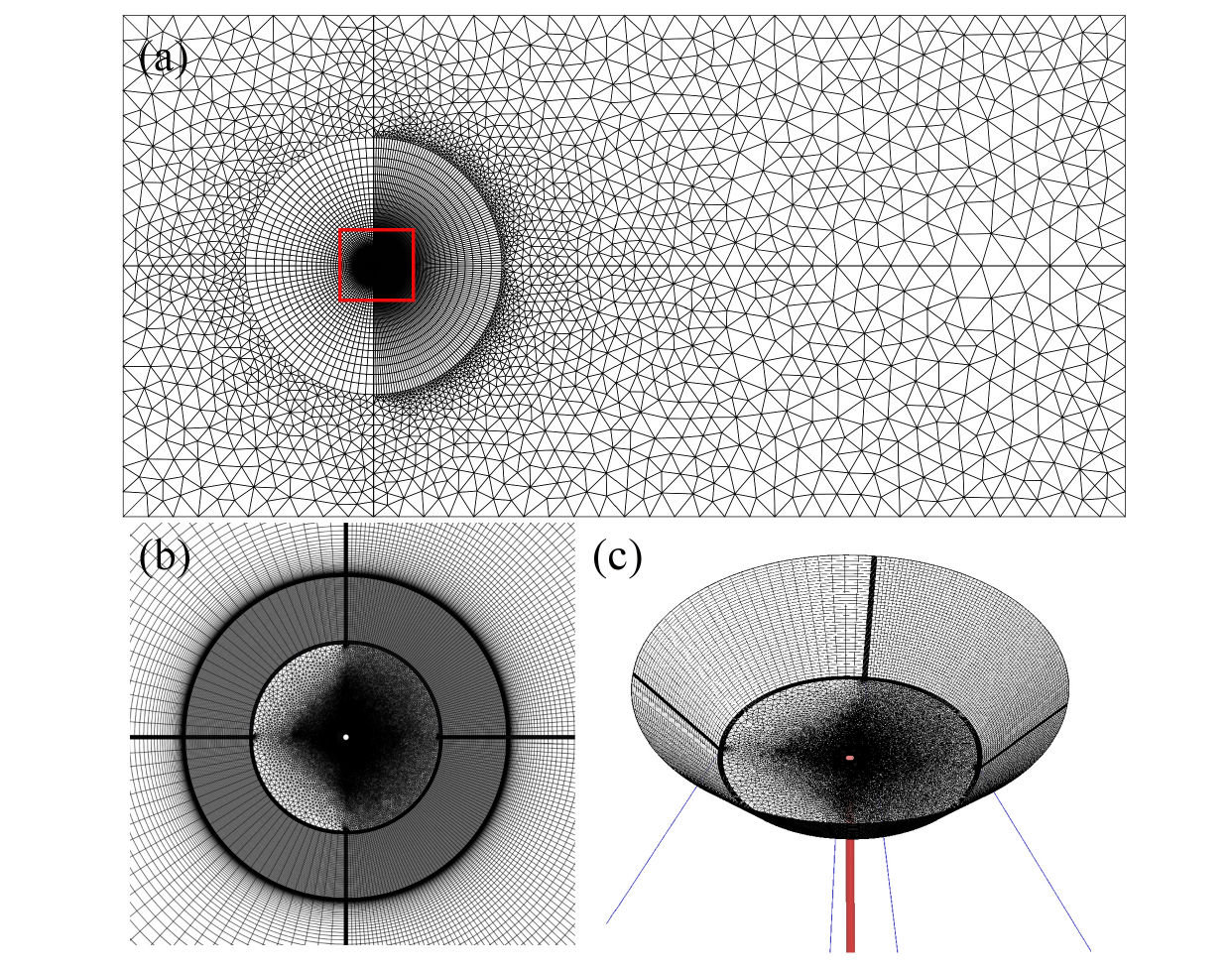}
	 \caption{Finite element discretization of the computational domain: (a) top-view of the fluid domain mesh, (b) close-up view of the mesh around the floater (inset box in (a)) and (c) isometric view of the mesh consisting of the floater, riser and the moorings.}
	 \label{fmr_mesh}
\end{figure}

A computational domain of size $[40D_f\times 20D_f\times 501.5D_r]$ has been considered for this study. The left hand side of the computational domain represents the inlet boundary and a uniform inflow enters the computational domain. The center of the floater is positioned $10D_f$ from the inlet boundary. The side walls of the computational domain are placed 10$D_f$ from the floater center. The top surface of the computational domain is considered as free-surface where slip boundary is implemented, i.e., $\sigma_{xy}=0$, $\sigma_{zy}=0$, $\bar{v}_y= 0$ and $ \frac{\partial\bar{\nu}}{\partial n}=0$, where $n$ is outward surface normal. A no-slip boundary condition ($\bar{v}_{x}= \bar{v}_{y} = \bar{v}_{z} = \bar{\nu} = 0$) is applied at the bottom of the computational domain which is considered as sea-bed. A traction-free boundary condition ($\sigma_{xx}=\sigma_{yx} = \sigma_{zx} = \frac{\partial \bar{\nu}}{\partial n}=0 $) is implemented at the outlet which is placed $30D_f$ away from the center of the floater. A no-slip boundary condition is implemented at the surface  of each component of the multibody system, i.e., floater, moorings and riser. 

The fluid domain is discretized into approximately 5 million nodes with a hybrid unstructured finite element mesh of 6-node wedge and 8-node brick elements. Each of the flexible line components, i.e., the riser and the moorings are discretized with 100 nonlinear beam and cable elements respectively. The requirement of the mesh for obtaining an economical solution for the problem of this scale is a challenge. Boundary layer and wake region in the $X$-$Y$ plane are resolved sufficiently and a relatively coarser mesh is used for the rest of the domain. Figure \ref{fmr_mesh} depicts the typical mesh used in the current study. The discretization in the direction parallel to the riser axis, i.e, the number of span-wise layers is taken as 100. A close-up view of the discretization around the floater is shown in Figs. \ref{fmr_mesh}b and \ref{fmr_mesh}c. The boundary layer thickness for all the flexible bodies are taken as 0.25$D$ ($D$ is the diameter of the component e.g., moorings, riser) with the stretching ratio, $\Delta y_{j+1}/\Delta y_j$ of 1.15 and satisfying $y^+ < 1$. A non-dimensional time step ($\Delta t U/D_r$) of 0.1 is chosen for this case as well. The dimensionless parameters used in the simulation are given in Table \ref{floater:tab}.
\begin{table}[]
	\centering
	\caption{Dimensionless parameters used in floater-mooring-riser simulation}
		\begin{tabular}{@{}lc@{}}
			\hline
			Dimensionless parameters &value\\ \hline
			\hline
	\vspace{0.25cm}
	$Re = \frac{\rho^\mathrm{f} UD_r}{\mu^\mathrm{f}}$ & $ 4000$  \\
	Floater: & \\
	\vspace{0.25cm}
	$m^*\ = \frac{m}{V_f\rho^\mathrm{f}}$&$ 0.98$ \\
				Riser: & \\
	$K_B = \frac{EI}{\rho^\mathrm{f} U^2D^4}$ & $ 2.1158 \times 10^7$ \\
	\vspace{0.25cm}
	$m^*\ = \frac{m}{\frac{\pi}{4}D^2L\rho^\mathrm{f}}$ & $ 2.23$ \\
	Mooring: & \\
	$K_A = \frac{EA}{\rho^\mathrm{f} U^2D^2}$ & $ 1.5708 \times 10^{11}$ \\

	
	\vspace{0.25cm} $m^*\ = \frac{m}{\frac{\pi}{4}D^2L\rho^\mathrm{f}}$ &$ 8.0$ \\
	[0.1cm]
	
			\hline
		\end{tabular}\label{floater:tab}
	\end{table}
The variables in Table \ref{floater:tab} are symbolized as follows: $E$ is the Young's modulus, $I$ is the second moment of area of the cross section, $m$ is the mass of the individual components of floater-mooring-riser system, $V_f$ is the volume of displaced fluid by the floater.

\begin{figure}[!htbp]
	\centering
	\begin{subfigure}{0.5\textwidth}
		\centering
		\includegraphics[width=\linewidth]{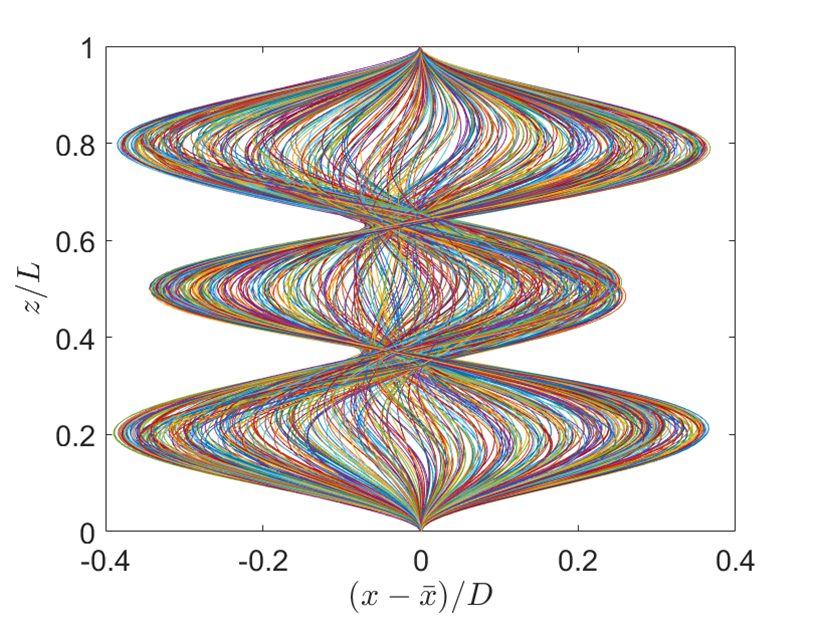}
		\caption{}
		\label{fig:floater_disp_x}
	\end{subfigure}
	\begin{subfigure}{0.48\textwidth}
		\centering
		\includegraphics[width=\linewidth]{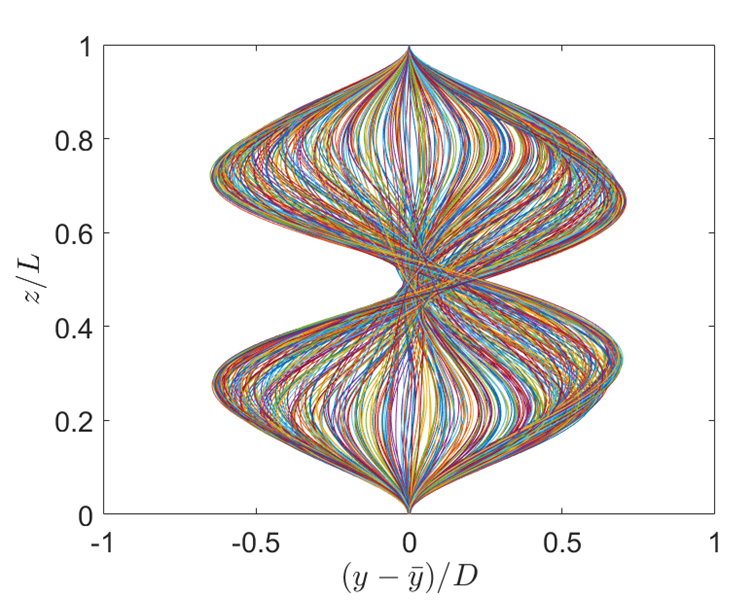}
		\caption{}
		\label{fig:floater_disp_y}
	\end{subfigure}
	\caption{Displacement response envelop for  riser component in floater-mooring-riser system subjected to uniform current flow: (a) cross-flow and (b) in-line directions. Higher modes, i.e., third mode in the in-line (IL) direction and a second mode in the cross-flow (CF) direction is observed in comparison to the pin-pin riser case.}
	\label{fig:response}
\end{figure}

\begin{figure}[!htbp]
	\centering
	\begin{subfigure}{0.5\textwidth}
		\centering
		\includegraphics[width=\linewidth]{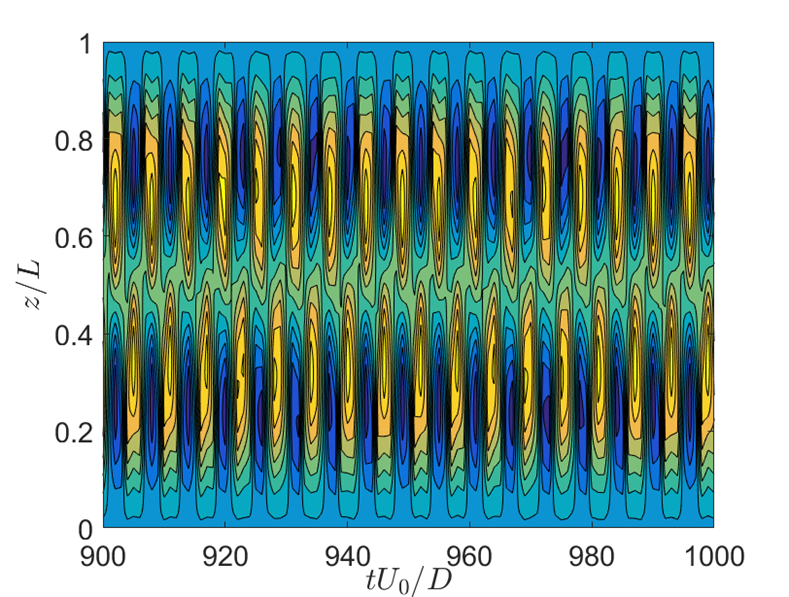}
		\caption{}
		\label{fig:twave_x}
	\end{subfigure}
	\begin{subfigure}{0.48\textwidth}
		\centering
		\includegraphics[width=\linewidth]{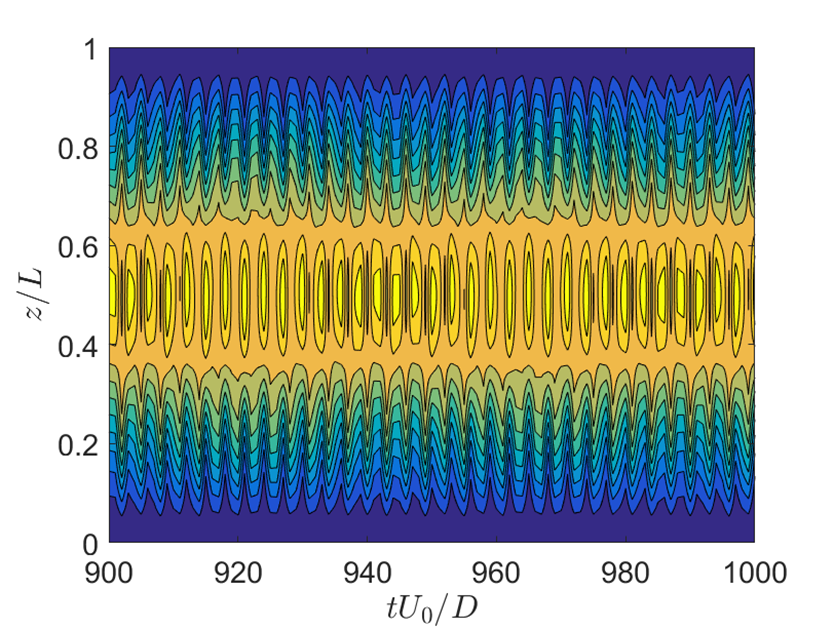}
		\caption{}
		\label{fig:twave_y}
	\end{subfigure}
	\caption{Standing wave response of the flexible riser component in floater-mooring-riser system subjected to uniform current flow: (a) cross-flow and (b) in-line directions.}
	\label{fig:twave}
\end{figure}

In Fig. \ref{fig:response}, the displacement response envelop of the riser is shown. It can be inferred that the riser vibrates with a dominant third mode in the in-line (IL) direction and a second mode in the cross-flow (CF) direction. In contrast, for the riser validation case (provided in Sec. \ref{Val})  lower modes are observed, i.e., second mode in the in-line (IL) direction and a first mode in the cross-flow (CF) direction (see Fig . \ref{riser:rms}). Also, relatively higher amplitudes in both in-line and cross-flow are observed in comparison to that of riser validation case. One of the possible reasons which can be attributed to this difference is the floater motion induced vibration. Even though the floater motion is small, i.e., $\mathcal{O}(10^{-3})$, it has a significant impact on the dynamics of riser. In Fig. \ref{fig:twave}, the riser response along the span is plotted as a function of time. A standing wave pattern is observed for both in-line and cross-flow responses of the riser. 

\begin{figure}[!htbp]
	\centering
	\begin{subfigure}{0.65\textwidth}
		\centering
		\includegraphics[width=\linewidth]{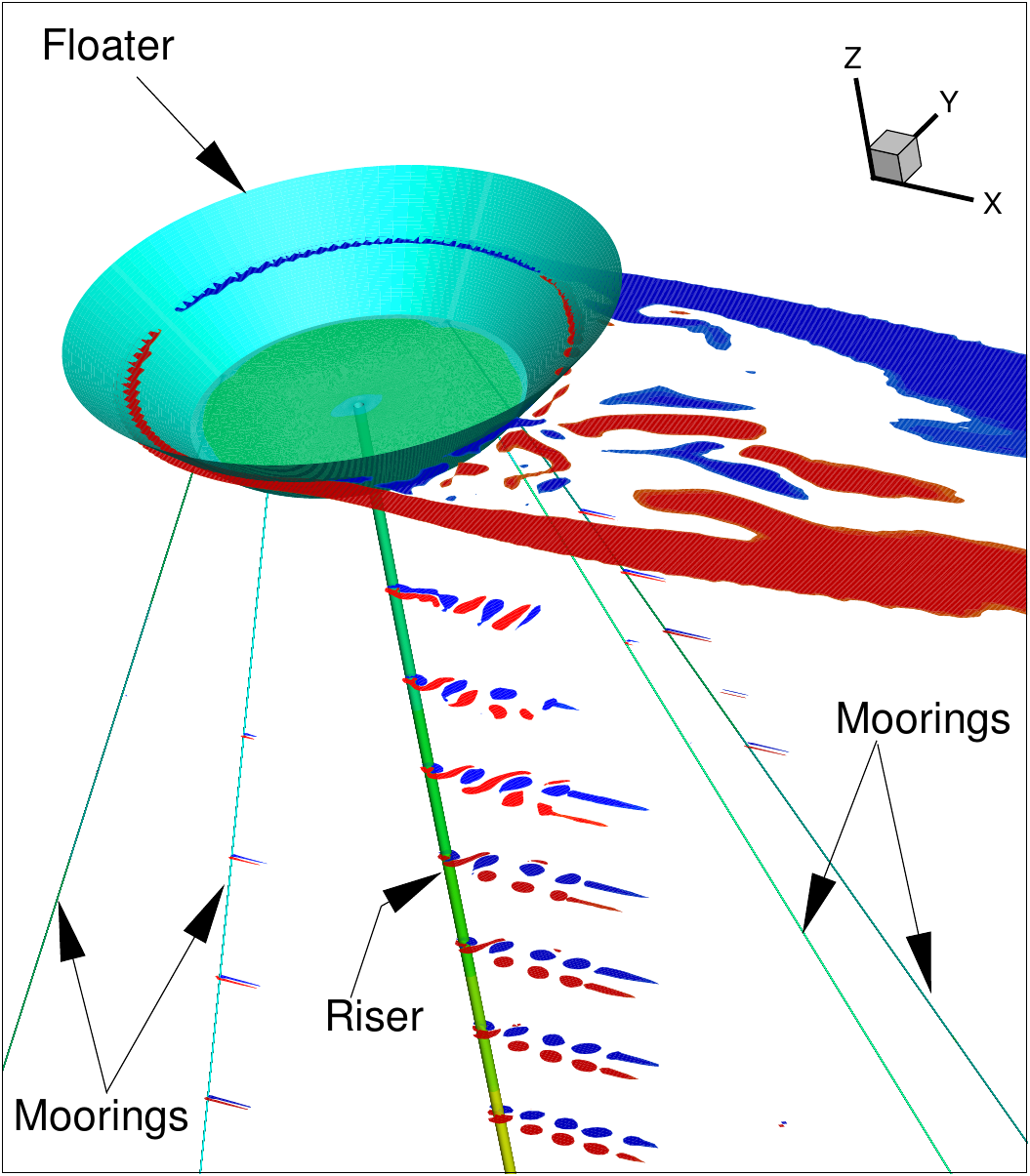}
	\end{subfigure}
	\caption{Instantaneous $Z$-vorticity behind the floater and the riser systems for the fully-coupled floater-mooring-riser system in a uniform current flow.}
	\label{fig:vort}
\end{figure}
Fig. \ref{fig:vort} shows a snapshot of the spanwise $Z$-vorticity for the floater-mooring-riser system. We observe a predominant 2S mode of vortex shedding pattern along most of the locations on the riser. Due to the large dimension of the floater, large vortices are formed along the surface of the floater. These large vortices either merge or destroy the small vortices formed by the riser and the moorings. Therefore, the proposed coupled fluid-flexible multibody solver has the capability to capture the physics of flow-induced vibration of the floater-mooring-riser system and it has significant implications on various offshore engineering applications.  
The present results successfully demonstrate the functionality and usability of the surface-to-line coupling method for a flexible multibody system interacting with a complex flow dynamics.
Finally, further investigations for various environmental conditions and floater-mooring-riser arrangements should be explored with the present framework.

\section{Conclusions}\label{Conclusion}
In the present manuscript, a general purpose partitioned iterative scheme is developed to integrate an incompressible 
turbulent fluid flow with a flexible multibody system. In particular, we have proposed a novel conservative 
surface-to-line projection procedure across nonmatching meshes between the 3D fluid flow 
and the 1D flexible structural bodies. 
Of particular interest to the offshore application, the present surface-to-line coupling procedure  
has been demonstrated for the interaction 
of turbulent current flow with mooring lines and marine riser modeled as nonlinear cables and beam, respectively. 
To achieve a stable and robust partitioned staggered coupling, 
the  nonlinear iterative force correction has been employed for the integration of 
rigid and flexible bodies with the incompressible turbulent flow.
Furthermore, a monolithic energy decaying scheme is used for solving the multibody system with constraints 
which makes the coupled fluid-flexible multibody solver efficient in solving problems where high frequencies 
are present and the multibody system is physically stiff.
The accuracy of the proposed method is validated by comparing the response characteristics of a flexible long riser 
 against the available experimental data. It is observed that the proposed line-to-surface or vice-versa coupling technique based on the conservative data transfer and the nonlinear iterative force correction have predicted the results with a practical accuracy for offshore engineering problems involving complex fluid-structure interaction and turbulence effects. Finally, the applicability of the present surface-to-line coupling method is demonstrated by simulating the fluid-structure interaction of a realistic offshore floater-mooring-riser system.
One of the natural extension of this work is to include the combined wave-current effects for the coupled dynamics of floating structures connected with risers and mooring lines.
While such fully-coupled fluid-flexible multibody simulation is computationally expensive, another possible extension of the present work may include the utilization of the high-fidelity data to construct a data-driven computing method. Such combination of high-fidelity data and the data-driven procedure will allow to explore a broad range of environmental conditions 
and  the real-time control of the multibody vibrations. 
   
\section*{Acknowledgments}
The authors would like to thank the National Research Foundation through Keppel-NUS Corporate Laboratory. The conclusions put forward reflect the views of the authors alone, and not necessarily those of the institutions. 
We also acknowledge the computing support  from the National Supercomputing Center Singapore.

\bibliographystyle{unsrt}
\bibliography{References}
\end{document}